\documentclass[aps,pre,10pt,showpacs,floatfix,onecolumn,nofootinbib,a4paper]{revtex4-2}
\usepackage{amssymb, amsmath, amsthm, mathtools}
\usepackage{bm}
\usepackage{mathrsfs}
\usepackage{hyperref,bookmark}
\usepackage{graphicx}
\usepackage{epsfig,color}
\usepackage{mathrsfs}
\usepackage{verbatim}
\usepackage{url}
\usepackage{subcaption}
\usepackage{float}
\usepackage[greek,english]{babel}
\usepackage{dcolumn}

\newcommand{\dd}{\operatorname{d}\!}

\newcommand{\nigh}[1]{{\color{black}{#1}}}

\newcommand{\diver}{\operatorname{div}}
\newcommand{\curl}{\operatorname{curl}}
\newcommand{\arccot}{\operatorname{arccot}}

\newcommand{\n}{\bm{n}}
\newcommand{\e}{\bm{e}}

\newcommand{\normal}{\bm{\nu}}
\newcommand{\body}{\mathscr{B}}
\newcommand{\free}{\mathscr{F}}
\newcommand{\core}{\mathscr{C}}

\newcommand{\Req}{R_\mathrm{e}}
\newcommand{\phic}{\phi_\mathrm{c}}

\newcommand{\framenn}{(\n_1,\n_2,\n)}

\newcommand{\conf}{\mathsf{S}}

\newcommand{\vae}{\varepsilon}

\newcommand{\alphab}{\alpha_\mathrm{b}}
\newcommand{\alphadown}{\alpha_\ast}
\newcommand{\alphaup}{\alpha^\ast}

\newcommand{\WOF}{W_\mathrm{OF}}
\newcommand{\WQ}{W_\mathrm{Q}}
\newcommand{\Wa}{W_\mathrm{a}}
\newcommand{\Ws}{W_\mathrm{s}}
\newcommand{\bend}{\bm{b}}

\newcommand{\Wn}{\mathbf{W}(\n)}
\newcommand{\Pn}{\mathbf{P}(\n)}
\newcommand{\Dn}{\mathbf{D}}
\newcommand{\I}{\mathbf{I}}

\newcommand{\distoc}{(S,T,b_1,b_2,q)}

\newcommand{\frameca}{(\e_x,\e_y,\e_z)}
\newcommand{\framexy}{(\e_x,\e_y)}

\newcommand{\region}{\mathscr{R}}
\newcommand{\boundaryR}{\partial\region}

\begin{document}
	\latintext
	
	\title{A geometric method to determine chromonics' planar anchoring strength}
	\author{Silvia Paparini}
	\email{silvia.paparini@unipv.it}
	\author{Epifanio G. Virga}
	\email{eg.virga@unipv.it}
	\affiliation{Department of Mathematics, University of Pavia, Via Ferrata 5, 27100 Pavia, Italy}
	\begin{abstract}
		Chromonic nematics are lyotropic liquid crystals that have already been known for half a century, but   have only recently raised interest for their potential applications in life sciences. Determining elastic constants and anchoring strengths for rigid substrates has thus become a priority in the characterization of these materials. Here, we present a method to determine chromonics' planar anchoring strength. We call it \emph{geometric} as it is based on recognition and fitting of the stable equilibrium shapes of droplets surrounded by the isotropic phase in a thin cell with plates enforcing parallel alignments of the nematic director. We apply our method to shapes observed in experiments; they resemble elongated rods with round ends, which are called \emph{b\^atonnets}. Our theory also predicts other droplets' equilibrium shapes, which are either slender and round,  called \emph{discoids}, or slender and pointed, called \emph{tactoids}. In particular, sufficiently small droplets are expected to display shape bistability, with two equilibrium shapes, one tactoid and one discoid, exchanging roles as stable and metastable shapes upon varying their common area.   
	\end{abstract}
	\date{\today}
	
	\maketitle
	
	\section{Introduction}\label{sec:intro}
	Liquid crystals come in two fashions: thermotropic and lyotropic. The former are condensed by reducing temperature, the latter by increasing concentration. Chromonic liquid crystals (CLCs) are lyotropic; they are formed by certain dyes, drugs, and nucleic acids. When added to water, these microscopic compounds assemble into \emph{stacks} of molecules, giving rise to rod-like structures that constitute the anisotropic components of the material. At appropriate temperatures and concentrations, these constituents form a nematic phase, which possesses only orientational order, or a more complex columnar phase, which also exhibits a certain degree of positional order \cite{lydon:chromonic,lydon:handbook,lydon:chromonic_1998,lydon:chromonic_2010,dierking:novel}. Here we shall only be concerned with the nematic phase of CLCs.
	
	Recent times have seen a surge of interest in these phases, mainly because they are soluble in water, and so promise to have valuable applications in life sciences \cite{park:lyotropic}. Indeed, success has already been granted to the use of CLCs to detect the presence of toxins and cancer biomarkes in simple devices \cite{shiyanovskii:real-time,woolverton:liquid,shaban:label-free}.
	
	These and other applications rely on a proper characterization of CLCs, including the determination of elastic constants and anchoring strength on rigid substrates. Here, we are especially interested in the latter. We shall propose a method to determine the strength of planar anchoring for chromonics; we call it \emph{geometric}, as it is based on shape recognition and fitting.
	
	The primary motivation for our study  came from the experiment performed in \cite{yi:orientation} on two-dimensional bipolar droplets of chromonic liquid crystals (CLCs) in the nematic phase at equilibrium with their isotropic phase,  placed  between two parallel plates inducing uniform director alignment. The bounding plates were patterned with sub-micron scale linear channels (equally aligned on both plates). 
	
	At sufficiently low CLC concentrations, the surface pattern resulted in a preferential alignment of the supramolecular stacks constituting the material with their axes along the surface channels. For large enough concentrations, however, the preferential alignment was still seen to lie on the bounding plates, but at right angles with the surface channels, a phenomenon confirmed and further analyzed in \cite{mcguire:orthogonal}. Here, we shall focus attention on the concentration regime for which channels tend to align stacks along their axis. Substrates with this property will be called \emph{aligning}. 
	
	Experiments showed essentially \emph{two-dimensional} droplets  bearing a \emph{bipolar} director field $\n$  with only in-plane components. It is precisely the stable equilibrium shape of these droplets that will be used to determine the anchoring strength of the bounding plates.
	
	The paper is organized as follows. 
	In Sec.~\ref{sec:problem}, we  extend the model employed in \cite{paparini:shape} for \emph{degenerate} substrates, that is, substrates for which no easy axis is prescribed on the bounding plates. Our analysis will build on our previous work, but we shall thrive to present it in a self-contained manner, leaving out only details that can be easily retrieved. In Sec.~\ref{sec:method}, we present our method and apply it to an exemplary case taken from \cite{yi:orientation}. Special care is given in Sec.~\ref{sec:alpha_safeguard} to make sure that the droplets under study are neither too small nor too large to question the tangential anchoring of the nematic director at their isotropic interface, which is a prerequisite of our analysis. Section~\ref{sec:bistability} is devoted to illuminate a typical bistability phenomenon, which we predict to occur for sufficiently small droplets: two different types of shapes coexist at equilibrium, one stable and the other metastable, exchanging their roles at a critical value of the area. Finally, in Sec.~\ref{sec:conclusion}, we collect the conclusions of this study. Two Appendices contain further mathematical details needed to justify a few key passages in our development.
	
	\section{Free boundary problem}\label{sec:problem}
	In this section, we present our mathematical theory focused on solving a  free boundary problem within a specific class of shapes and director fields.
	\subsection{Energetics}\label{sec:energetic}
	We shall denote by $\n$ the director field which represents on a macroscopic scale the average orientation in space of the supramolecular stacks that constitute a CLC. A solid body of experimental evidence (see, for example, \cite{nayani:spontaneous,davidson:chiral,fu:spontaneous}) suggests that the ground state of CLCs is \emph{not} the one with $\n$ uniform in space, as customary in ordinary nematics, but a distorted one, often characterized as a \emph{double twist}.
	
	To appreciate this better, we recall that according to the decomposition of $\nabla\n$ proposed by \cite{machon:umbilic} and reprised and reinterpreted by \cite{selinger:interpretation} we can write
	\begin{equation}
		\label{eq:nabla_n_decomposition}
		\nabla\n=-\bend\otimes\n+\frac12T\Wn+\frac12S\Pn+\Dn,
	\end{equation} 
	where $S:=\diver\n$ is the \emph{splay}, $T:=\n\cdot\curl\n$ is the \emph{twist}, $\bend:=\n\times\curl\n$ is the \emph{bend}, $\Wn$ is the skew-symmetric tensor associated with the axial vector $\n$, $\Pn:\I-\n\otimes\n$ is the projection onto the plane orthogonal to $\n$, and $\Dn$ is a symmetric traceless tensor such that $\Dn\n=\bm{0}$, which can be represented as 
	\begin{equation}
		\label{eq:D_representation}
		\Dn=q(\n_1\otimes\n_1-\n_2\otimes\n_2).
	\end{equation}
	In \eqref{eq:D_representation}, $(\n_1,\n_2)$ is an orthonormal pair in the plane orthogonal to $\n$ and $q\geqq0$ is a scalar measure of distortion that we call \emph{octupolar splay} \cite{pedrini:liquid}. The bend vector $\bend$ can be decomposed in in the frame $\framenn$ as $\bend=b_1\n_1+b_2\n_2$. We call the scalars $\distoc$ distortion \emph{characteristics} of a director field $\n$.
	
	A \emph{single} twist is a state of distortion for which $T=\pm2q$, with $q$ constant and all other distortion characteristics zero; this is, for example, the ground state of cholesterics \cite{selinger:director}. On the other hand, a \emph{double} twist is a state of distortion for which $T$ is the only constant distortion characteristic that does not vanish. What distinguishes a double twist from a single twist is that the latter can fill three-dimensional space, whereas the former cannot \cite{virga:uniform}. Thus, a double twist in the ground state turns necessarily into a source of \emph{frustration}, as it can be achieved locally, but not globally. 
	
	The appropriate form of the elastic free energy for CLCs has lately become a matter of debate. The classical Oseen-Frank form $\WOF$, written as \cite{selinger:interpretation}
	\begin{equation}
		\label{eq:Oseen_Frank_energy}
		\WOF=\frac12(K_{11}-K_{24})S^2+\frac12(K_{22}-K_{24})T^2+\frac12K_{33}B^2+2K_{24}q^2,
	\end{equation}
	where $B^2:=\bend\cdot\bend$ and $K_{11}$, $K_{22}$, $K_{33}$, and $K_{24}$ are the \emph{splay}, \emph{twist}, \emph{bend}, and \emph{saddle-splay} constants, has been employed with
	\begin{equation}
		\label{eq:violated_inequality}
		K_{24}>K_{22},
	\end{equation}
	which aims at making a double twist distortion energetically preferred to no distortion, in violation of one Ericksen's inequality \cite{ericksen:inequalities}.  These inequalities, which are immediately read off from \eqref{eq:Oseen_Frank_energy}, would instead ensure that $\WOF$ is positive semi-definite: they are
	\begin{equation}
		\label{eq:Ericksen_inequalities}
		K_{11}\geqq K_{24}\geqq0,\quad K_{22}\geqq K_{24}\geqq0,\quad\text{and}\quad K_{33}\geqq0.	
	\end{equation}
	It has, however, been shown in \cite{paparini:paradoxes} that paradoxical consequences follow from \eqref{eq:violated_inequality} for the equilibrium shape of CLC droplets  surrounded in three space dimensions by an isotropic fluid, although for fixed domains a variational theory based on \eqref{eq:Oseen_Frank_energy} could be viable \cite{paparini:stability}. Other perplexing consequences of \eqref{eq:violated_inequality} were also found for fixed domains in \cite{long:violation}.
	
	To remedy such a state of affairs and have a theory that could be applied to fixed and variable domains alike, we proposed in \cite{paparini:elastic} a quartic twist elastic theory based on a density $\WQ$ that differs from $\WOF$ by a single term proportional to $T^4$. Further consequences of this theory were also examined in \cite{paparini:spiralling}. 
	
	Here we do not delve any longer on this possibly controversial issue, as in the setting to which our theory will be applied the director field $\n$ is \emph{planar}, that is, it lies everywhere parallel to a given plane and is independent of the coordinate orthogonal to that plane. For such a field (see also \cite{pedrini:relieving}),
	\begin{equation}
		\label{eq:characteristics_planar}
		T=0\quad\text{and}\quad S^2=4q^2,
	\end{equation}  
	and so \emph{both} $\WOF$ and $\WQ$ reduce to 
	\begin{equation}
		\label{eq:planar_reduced_energy}
		W=\frac12K_{11}S^2+\frac12K_{33}B^2,
	\end{equation} 
	which is a well-behaved positive-definite energy density for
	\begin{equation}
		\label{eq:reduced_inequalities}
		K_{11}>0\quad\text{and}\quad K_{33}>0,
	\end{equation}
	the only inequalities needed below. 
	
	Where a CLC is in contact with its coexistent isotropic phase, an anisotropic surface tension is present along the interface, which we shall represent by the classical Rapini-Papoular formula \cite{rapini:distortion},
	\begin{equation}
		\label{eq:Rapini_Papoular_formula}
		\Ws=\gamma[1+\omega(\n\cdot\normal)^2],
	\end{equation}
	where $\gamma>0$ is the isotropic component of the surface tension at the interface, $\omega>-1$ is a dimensionless parameter weighting the anisotropic component, and  $\normal$ is a unit normal to the interface. Here, $\omega$ is assumed to be positive,  so that the interfacial energy density (per unit area) $\Ws$ is minimized when $\n$ is tangent to the interface. Where a CLC is in contact with an aligning substrate, the anchoring energy density $\Wa$ is represented similarly to \eqref{eq:Rapini_Papoular_formula} as
	\begin{equation}
		\label{eq:anchoring_energy}
		\Wa=\frac12\sigma_0[1-(\n\cdot\e)^2],
	\end{equation}
	where $\sigma_0>0$ is the \emph{anchoring strength} and $\e$ is a unit vector designating the \emph{easy} axis. In \eqref{eq:anchoring_energy}, $\Wa$ is normalized so as to have zero minimum.
	
	\subsection{2D Variational Problem}\label{sec:variational}
	Our theory aims to explain the experiment conducted in \cite{yi:orientation}, which involved coexisting  nematic and isotropic phases of a CLC confined within two parallel plates, both treated so as to induce one and the same parallel uniform easy axis on the nematic phase.\footnote{Alignment was achieved by use of topographic patterns that had already been  characterized for thermotropic liquid crystals \cite{yoon:organization,kim:alignment,behdani:alignment}.}
	
	The experimental setting suggests a few assumptions that we shall adopt in our analysis. First, the director field $\n$ in the nematic phase is parallel to the bounding plates throughout the cell (and so it is a  planar field). Second, each nematic region to which we confine attention will be considered as a cylindrical island $\body$ of prescribed volume $V_0$ occupying the whole gap between the bounding plates. Third, the cross-section $\region$ of $\body$ will be symmetric about the easy axis $\e$. We now formulate in a mathematical language these hypotheses.
	
\begin{figure}[h] 
		\begin{subfigure}[c]{0.33\linewidth}
			\centering
			\includegraphics[width=.85\linewidth]{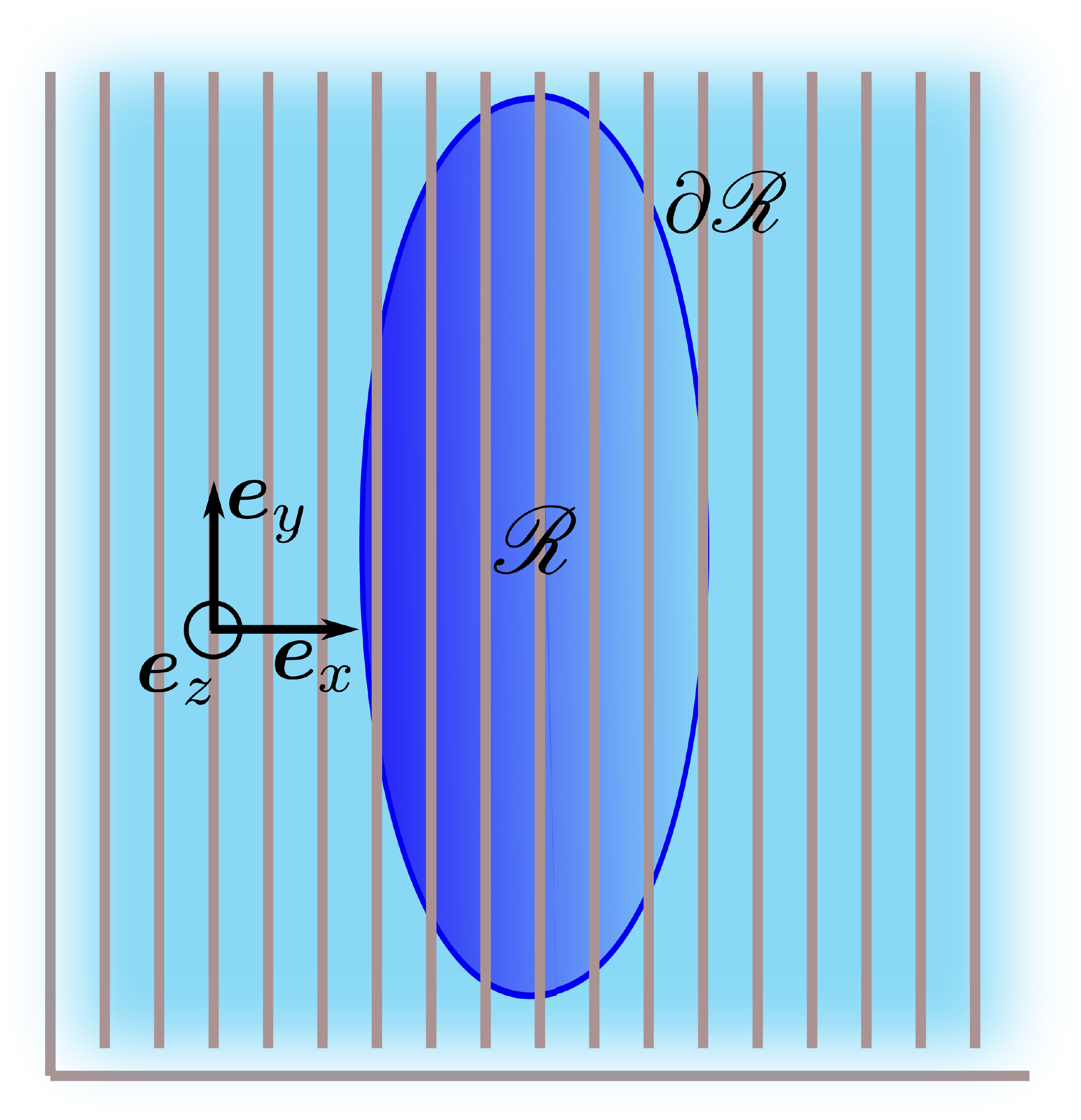}
			\caption{The region $\region$ superimposed to the (schematic) orienting sub-micron channels.}
			\label{fig:shape_clark_a}
		\end{subfigure}
		\begin{subfigure}[c]{0.33\linewidth}
			\centering
			\includegraphics[width=.5\linewidth]{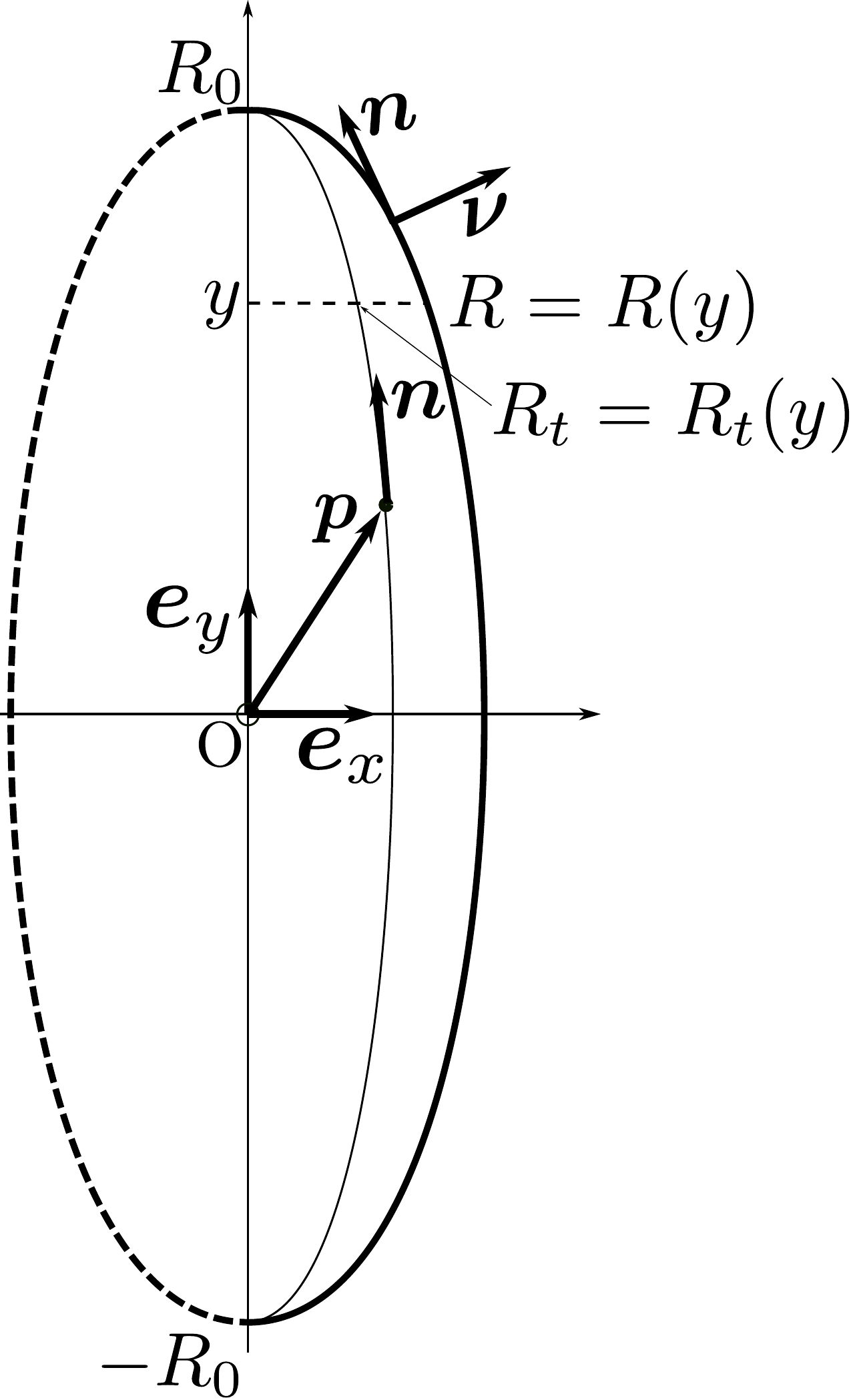}
			\caption{The boundary $\partial\region$ of $\region$ with outer unit normal $\normal$.}
			\label{fig:shape_clark_b}
		\end{subfigure}
		\caption{(Color online) Two-dimensional problem designed to reproduce the experimental setting of \cite{yi:orientation}. A CLC island $\body$ surrounded by the isotropic phase is squeezed between two parallel plates patterned with sub-micron line channels oriented along $\e_y$. The cross-section $\region$ of the island has area $A_0$ and it is taken to be mirror symmetric about both $x$ and the $y$ axes. Half of the boundary $\partial\region$  is represented as graph of a smooth, even function $x=R(y)$, where $y\in[-R_0,R_0]$. The bipolar director field $\n$ lies on the same plane as $\region$ and is defined as the unit tangent field to the family of curves $x=R_t(y)=g(t)R(y)$, which represent the retractions of $\partial\region$ inside $\region$ for generic $t\in[0,1]$. Both $\region$ and $\n$ should be thought of as uniformly extended through the gap between the parallel plates that bound the cell, at distance $h$ from one another, both orthogonal to the $z$ axis (which comes out of the plane of the figure).}
		\label{fig:shape_clark}
	\end{figure}

	We take the aligning bounding plates parallel to the $(x,y)$ plane of a Cartesian frame $\frameca$, with $\e_y$ coincident with easy axis $\e$. We  represent the island $\body$ as $\region\times[-\frac{h}{2},\frac{h}{2}]$, where $\region$ is a region with piecewise smooth boundary $\partial\region$ in the $(x,y)$ plane (see Fig.~\ref{fig:shape_clark_a}) and $h$ is the cell's thickness. 
	The  director field $\n$ lies everywhere in the $(x,y)$ plane and is independent of $z$. 
	
	The isoperimetric constraint on the volume of $\body$ translates into a constraint on the area of $\region$,
	\begin{equation}
		\label{eq:area_constraintA0}
		A(\region)=A_0,
	\end{equation}
	where $A$ is the area measure and $A_0=V_0/h$.
	
	With $\omega>0$ in \eqref{eq:Rapini_Papoular_formula}, the interfacial energy density $\Ws$ is minimized when $\n$ is tangent to $\partial\region$. For simplicity, we shall assume that 
	\begin{equation}
		\label{eq:degenerate_boundary_condition_2d}
		\n\cdot\normal\equiv0, \quad \hbox{on} \quad \boundaryR,
	\end{equation}
	and we shall treat it as a constraint on $\n$, save checking its validity \emph{a posteriori} with appropriate energy comparisons (see Sec.~\ref{sec:alpha_safeguard}) to  ensure that such a \emph{tangential anchoring} is not broken. In particular, we expect that for  aligning substrates assumption \eqref{eq:degenerate_boundary_condition_2d} may fail to hold for both sufficiently small droplets, as was shown to be  the case for degenerate substrates  \cite{paparini:shape}, and for sufficiently large droplets, for which the anchoring energy $\Wa$ is more likely to prevail over $\Ws$. 
	
	Adding all energy contributions discussed above, with the aid of \eqref{eq:planar_reduced_energy}, \eqref{eq:Rapini_Papoular_formula}, and \eqref{eq:anchoring_energy}, we arrive at the following total free-energy functional $\free$, which describes a chromonic island $\body$ with cross-section $\region$,
	\begin{equation}
		\label{eq:free_energy_functional_aligning}
		\free[\region;\n]:=h\left\{\frac12\int_{\region}[K_{11}(\diver\n)^2+K_{33}|\n\times\curl\n|^2]\dd A+\gamma \ell(\boundaryR)\right\}-\sigma_0\int_{\region} \left(\n\cdot\e_y\right)^2 \dd A,
	\end{equation}
	where  $\ell(\boundaryR)$ denotes the length of $\boundaryR$ and use has been made of \eqref{eq:degenerate_boundary_condition_2d}. In \eqref{eq:free_energy_functional_aligning}, an additive constant, $\sigma_0A_0$, has been omitted; it plays no role in the minimum problem studied here, in force of the isoperimetric constraint \eqref{eq:area_constraintA0}, which prescribes the area $A_0$ of the admissible domains $\region$.
	
	\subsection{Shapes and Director Fields}\label{sec:admissible_shapes}
	The shape of $\region$ is the primary unknown of our minimum problem. Free boundary problems like this are usually very difficult when treated in great generality, even in two space dimensions (as also witnessed by a recent analytic study \cite{geng:two-dimensional}). Following a well-established tradition (see, for example, the papers \cite{kaznacheev:nature,kaznacheev:influence,prinsen:shape,prinsen:parity,prinsen:continuous,puech:nematic,verhoeff:tactoids}), we shall tackle this problem in a special class of admissible shapes for $\region$.
	
	Inspired by the experimental setting of~\cite{yi:orientation}, we shall assume that the long axis of $\region$ is aligned with $\e_y$ and that the shape of $\region$ is mirror symmetric about both the $x$ and the $y$ axes  of the frame $\framexy$. Thus, only half of the curve that bounds $\region$ needs to be described, the other half being obtained by mirror symmetry. More precisely, $\boundaryR$ will be described as the graph of  a smooth, even function $x=R(y)$ defined over the interval $[-R_0,R_0]$, which is to be determined. The function $R$ vanishes at the end-points of this interval; they designate the \emph{poles} of the drop (see Fig.~\ref{fig:shape_clark_b}),
	\begin{equation}
		\label{eq:R_definition}
		R(\pm R_0)=0.
	\end{equation}
	Smoothness and symmetry require $R'(0)=0$, where a prime denotes differentiation with respect to the argument.
	Whenever $R'(R_0)$ is finite, the surface normal is discontinuous at the poles and the shape $\region$ is a \emph{genuine} tactoid; conversely, when $R'(R_0)$ is unbounded, $\boundaryR$ is  everywhere smooth.
	
	We shall use the method devised in \cite{paparini:shape} to obtain a bipolar director field $\n$ inside $\region$ from the mere knowledge of $\boundaryR$; this consists in retracting $\boundaryR$ inside $\region$ to generate  a family of  non-intersecting curves filling the whole of $\region$ with $\n$ everywhere tangent to them. More precisely, $\n$ is defined as the unit vector field tangent to the retracting inner curve $R_t(y):=g(t)R(y)$, where $t\in[0,1]$ and $g$ is an increasing function such that $g(0)=0$ and $g(1)=1$. Fig.~\ref{fig:shape_clark_b}  illustrates a sketch for such retracting lines; we will find the total free energy to be independent of the specific function $g$, and so of the specific method of retraction. 
	
	We rescale both $y$ and $R(y)$ to the radius $\Req$ of the \emph{equivalent} disc with area $A_0$, keeping their names unchanged, while we denote by $\mu$ the  ratio
	\begin{equation}
		\label{eq:mu_definition}
		\mu:=\frac{R_0}{\Req}.
	\end{equation}
	With this normalization, the area constraint \eqref{eq:area_constraintA0} reads simply as
	\begin{equation}
		\label{eq:area_rescaled}
		\int_{-\mu}^{\mu} R(y) \dd y=\frac\pi2
	\end{equation}
	and \eqref{eq:R_definition} becomes
	\begin{equation}\label{eq:end_point_condition}
		R(\pm\mu)=0.
	\end{equation}	
	Reasoning as in \cite{paparini:shape}, we can reduce $\free$ in \eqref{eq:free_energy_functional_aligning} to the following  functional in the scaled variables $y$ and $R(y)$,
	\begin{equation}
		\label{eq:F_energy_functional_aligning} 
		\mathcal{F}[\mu;R]:=\frac{\free[\region;\n]}{K_{11}h}=\mathcal{F}_\mathrm{e}[\mu;R]+\mathcal{F}_\mathrm{a}[\mu;R],
	\end{equation}
	where
	\begin{align}
		\label{eq:F_energy_functional} 
		\mathcal{F}_\mathrm{e}[\mu;R]:=\int_{-\mu}^{\mu}&\left\{\left[\frac{R'}{R}-\dfrac{R''}{R'}+\frac{1}{8}\dfrac{RR''^2}{R'^3}\left(3+k_3\right)\right]\arctan R'+\frac{R''}{1+R'^2}\right. \nonumber\\
		&\left.+\frac{1}{8}\frac{RR''^2}{(1+R'^2)^2}\left[(k_3-5)-\frac{1}{R'^2}(3+k_3)\right]+2\alpha\sqrt{1+R'^2}\right\}\dd y
	\end{align}
	is the scaled elastic free energy functional [see equation (11) of \cite{paparini:shape}], while  
	\begin{equation}
		\label{eq:free_sub_aligning_scaled}
		\mathcal{F}_\mathrm{a}[\mu;R]:=-4\beta\alpha^2\int_{-\mu}^{\mu}\frac{R}{R'}\arctan R'\dd y
	\end{equation}
	is the appropriate dimensionless form of the anchoring energy for aligning substrates (see Appendix~\ref{sec:retracted_field}).
	In \eqref{eq:F_energy_functional},
	\begin{equation}
		\label{eq:elastic_constants_rescaled}
		k_3:=\frac{K_{33}}{K_{11}}
	\end{equation}
	is the \emph{reduced bend} constant and 
	\begin{equation}
		\label{eq:alpha}
		\alpha:=\frac{\gamma\Req}{K_{11}}
	\end{equation}
	is a \emph{reduced area}.\footnote{Equivalently, $\alpha=\Req/\xi_\mathrm{e}$, where $\xi_\mathrm{e}$ is the de Gennes-Kleman \emph{extrapolation length} \cite[p.\,159]{kleman:soft}. In this language, a drop is either \emph{small} or \emph{large}, whether  $\alpha\ll1$ or $\alpha\gg1$, respectively.\label{foot:alpha}} In \eqref{eq:free_sub_aligning_scaled}, 
	\begin{equation}
		\label{eq:beta}
		\beta:=\frac{\sigma_0 K_{11}}{2h\gamma^2}
	\end{equation}
	is the  \emph{dimensionless anchoring} strength of the substrates.
	
	\nigh{Both $\alpha$ and $\beta$ are dimensionless parameters: while $\alpha$ is the ratio of two forces, $\beta$ has \emph{not} an equally transparent physical interpretation. The parameter $\widetilde{\beta}:=2\alpha^2\beta=\sigma_0\Req^2/K_{11}h$ would seem to be more meaningful, as it is the ratio of two energies. However, as will become clear in Sec.~\ref{sec:method}, in the present setting (for $\Req\sim10\,\mu\mathrm{m}$ and $\sigma_0\sim10\text{-}10^2\,\mu\mathrm{Jm^{-2}}$) we expect that $\alpha\sim10^2$ and $\widetilde{\beta}\sim10^2\text{-}10^3$, which makes $\beta\sim10^{-2}\text{-}10^{-1}$. Here, to simplify comparison with experiment, we choose to estimate $\alpha$ from other sources and not to use it as an independent fitting parameter. This makes $\beta$ and $\widetilde{\beta}$ interchangeable.
		
		According to \eqref{eq:F_energy_functional}, the value $k_3=5$ seems to be somewhat special: for $k_3>5$, two similar terms in the integrand would become antagonistic; we lack an explanation for this, but we heed that $k_3<5$ for the material of the experiment in \cite{yi:orientation} (see Sec.~\ref{sec:method}).}
	
	As a consequence of \eqref{eq:end_point_condition}, the functional $\mathcal{F}_\mathrm{e}$ in \eqref{eq:F_energy_functional_aligning} diverges logarithmically to $+\infty$ near the poles of $\region$, due to the non-integrablity at $y=\pm\mu$ of the integrand $\frac{R'}{R}\arctan R'$.\footnote{\nigh{Despite the apparent similarity between this integrand and that in \eqref{eq:free_sub_aligning_scaled}, the latter stays bounded, irrespective of the limiting value of $R'(y)$ for $y\to\pm\mu$ (no matter whether finite or not).}} The poles of $\region$ are also the points where $\n$ exhibits \emph{surface} defects, also known as \emph{boojums}; following a common practice (see, for example, \cite[p.\,171]{degennes:physics}) we tame these singularities by replacing them with isotropic \emph{cores} of size $\varepsilon$ (in $\Req$ units). Thus, the integrals in \eqref{eq:F_energy_functional_aligning}  and \eqref{eq:F_energy_functional} can be taken over a shorter interval $[-\eta,\eta]$, where $\eta$ is such that \cite{paparini:shape}
	\begin{equation}
		\label{eq:y_bar}
		R(\eta) = R (-\eta) = \varepsilon.
	\end{equation}
	For $\Req$ in the order of $10\,\mu\mathrm{m}$, it is reasonable to take $\vae\sim10^{-3}$. \nigh{As shown in Appendix~\ref{sec:core_boojum}, the extra energy stored in the defect cores does \emph{not} depend on $\varepsilon$ and is negligible compared with the surface anchoring energy, so that it  plays no role in the global minimum problem.} 
	
	\subsection{Special Family of Shapes}\label{sec:degenerate_substrates_family_shape}
\begin{figure}[h]
		\centering
		\begin{subfigure}[b]{0.2\linewidth}
			\centering
			\includegraphics[width=1.0\linewidth]{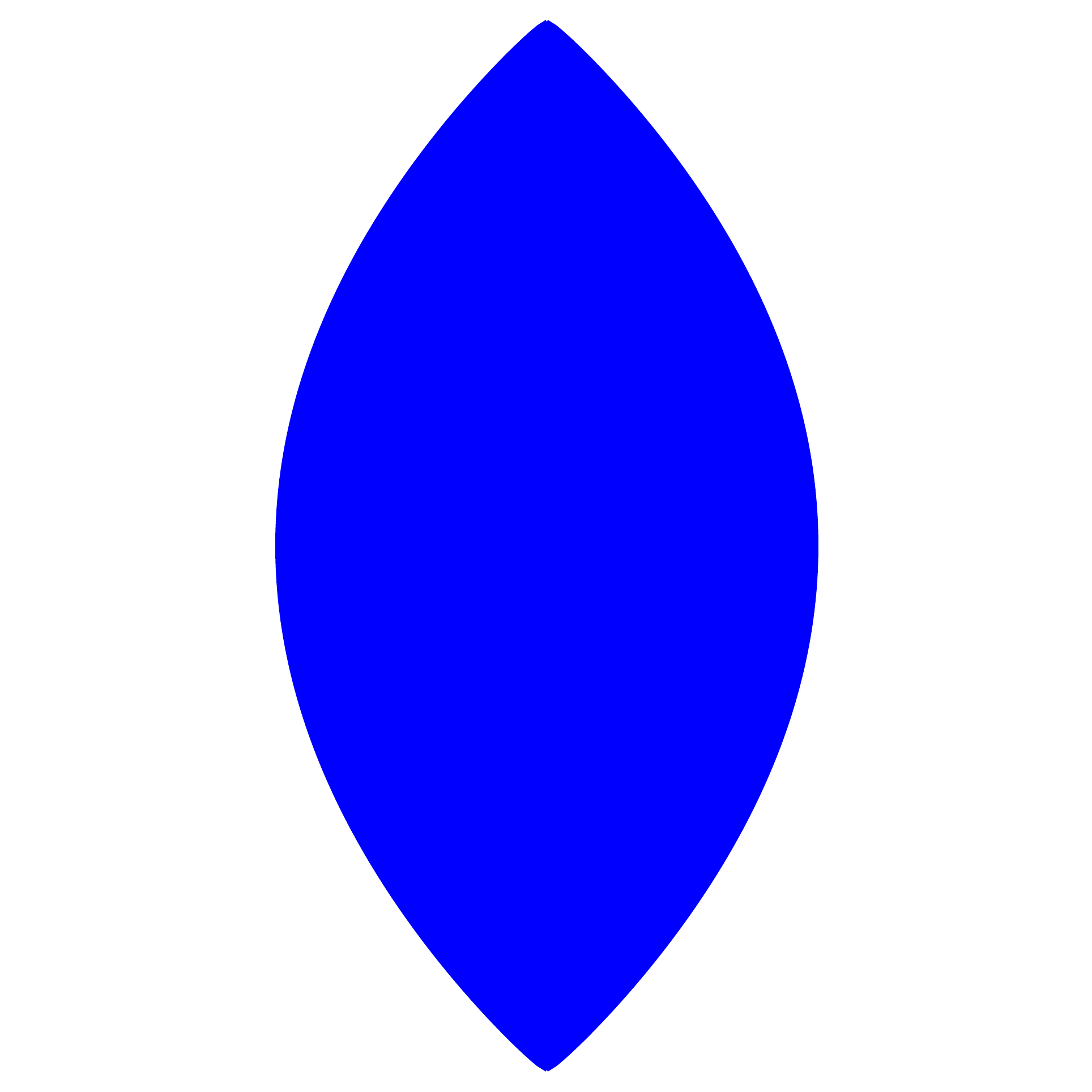}
			\caption{tactoid\qquad\qquad $\phi=0.1$\quad$\mu=1.5$}
			\label{fig:tacoid}
		\end{subfigure}
		\quad
		\begin{subfigure}[b]{0.2\linewidth}
			\centering
			\includegraphics[width=0.73\linewidth]{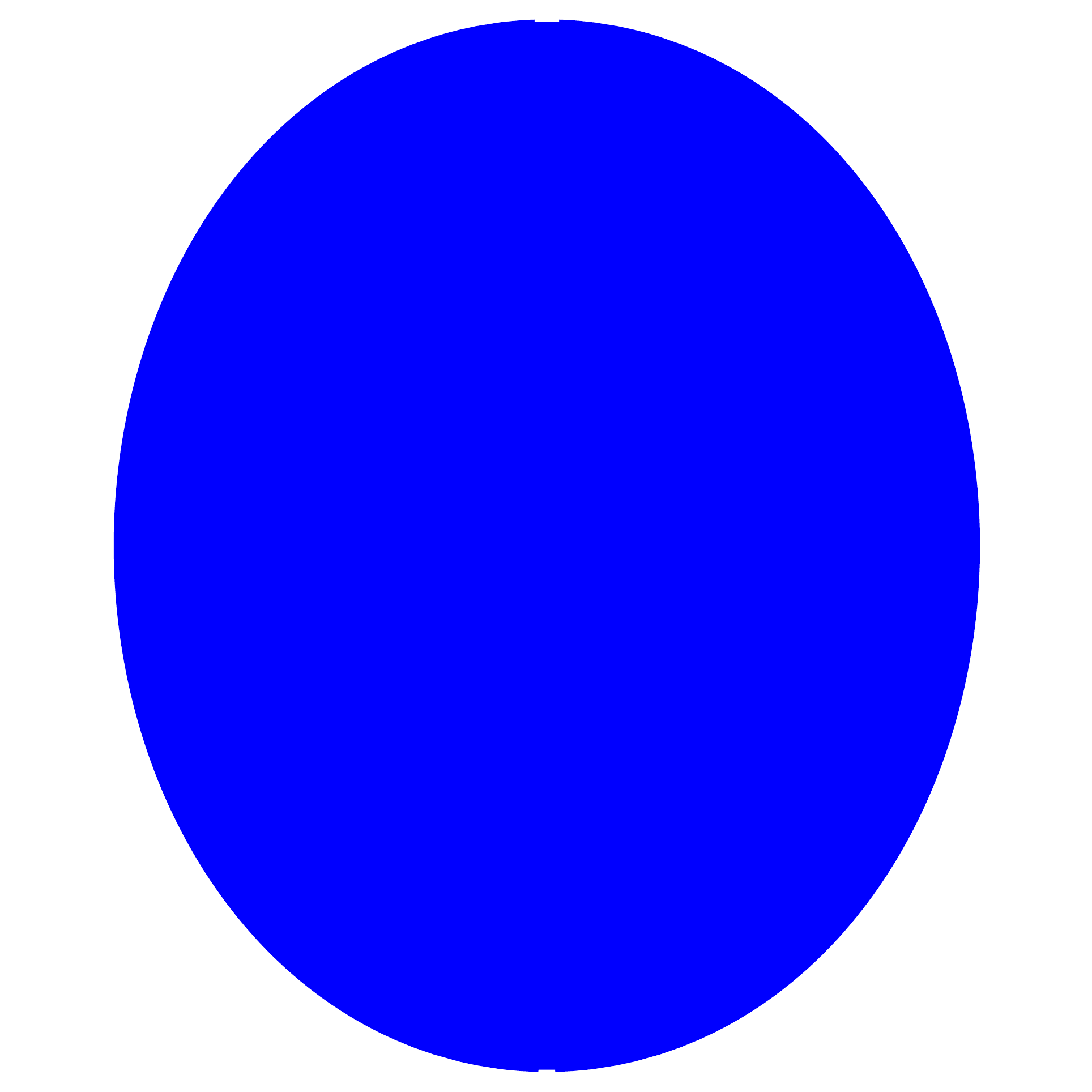}
			\caption{discoid\qquad\qquad $\phi=1.6$\quad$\mu=1.1$}
			\label{fig:discoid}
		\end{subfigure}
		\quad
		\begin{subfigure}[b]{0.2\linewidth}
			\centering
			\includegraphics[width=1.0\linewidth]{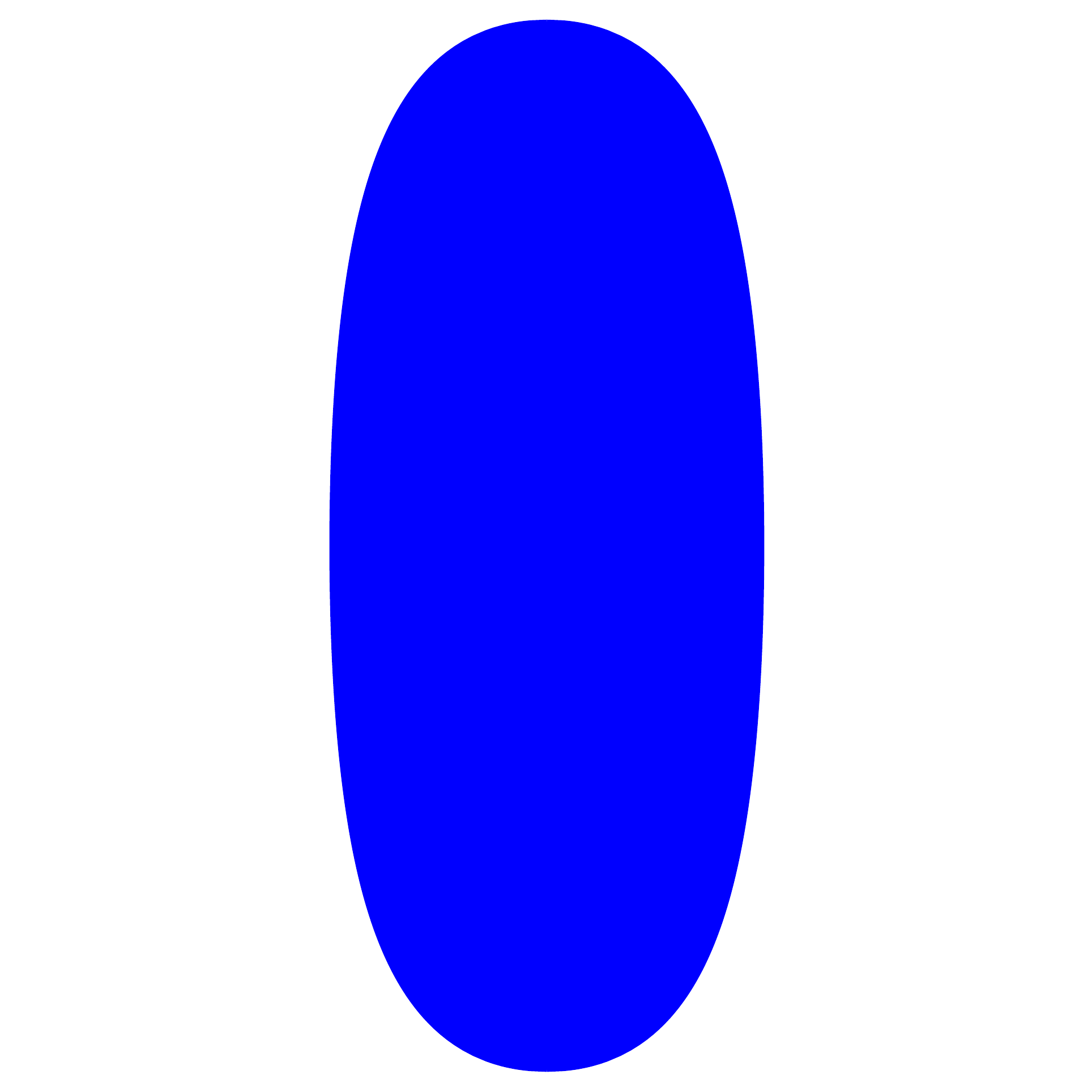}
			\caption{b\^{a}tonnet\qquad\qquad  $\phi=1.9$\quad$\mu=1.5$}
			\label{fig:batonnet}
		\end{subfigure}
		\caption{Gallery of exemplary shapes with profile as in \eqref{eq:profile} obtained for different values of parameters $(\phi,\mu)$, each showing a different type of possible convex minimizer of $\mathcal{F}$ in  \eqref{eq:F_energy_functional_aligning}. Tactoids  \emph{conventionally} occur for $0\leqq\phi<0.20$, discoids for $0.20\leqq\phi<1.77$, and b\^atonnets for $1.77\leqq\phi<2.03$. The value of $\phi$ assigns the type of shape represented by \eqref{eq:profile}, while $\mu$ affects its aspect ratio according to \eqref{eq:aspect_ratio}: for given $\phi$, the polar distance  grows quadratically with $\mu$ relative to the width.}
		\label{fig:gallery_aligning}
	\end{figure}

	Here, to simplify the analysis of the functional $\mathcal{F}$ in \eqref{eq:F_energy_functional_aligning}, we represent the admissible shapes of $\region$ by a two-parameter family of functions $R(y)$. Specifically, we let
	\begin{equation}
		\label{eq:profile}
		R(\phi,\mu;y)=\frac{\pi}{H(\phi)}\frac{1}{\mu}\left[\left(1-\left(\frac{y}{\mu}\right)^2\right)\cos\phi+\sqrt{1-\left(\frac{y}{\mu}\right)^2}\sin\phi\right],\quad\text{with}\quad H(\phi):=\frac{8}{3}\cos\phi+\pi \sin\phi>0.
	\end{equation}
	The parameters $(\phi,\mu)$ describe a two-dimensional configuration space $\conf:=\{(\phi,\mu):0\leqq\phi\leqq\frac{3\pi}{4}, \mu>0\}$, where the bounds on $\phi$ imply that $H>0$. More details on the geometric construction that justifies \eqref{eq:profile} can be found in \cite{paparini:shape}. Here, we only heed that while  by \eqref{eq:end_point_condition} $\mu$ represents the polar distance of $\region$, different values of $\phi$ affect its shape: conventionally, $\region$ is said to be a \emph{tactoid} for $0\leqq\phi<\frac{\pi}{16}$, a \emph{discoid} for $\frac{\pi}{16}\leqq\phi<\frac{9\pi}{16}$, and a \emph{b\^atonnet}\footnote{Name borrowed from French, meaning \emph{short staff}.} for $\frac{9\pi}{16}\leqq\phi<\phic:=\arccot(-\frac12)$. Genuine tactoids occur only for $\phi=0$. Representatives of these types of admissible shapes are depicted in Fig.~\ref{fig:gallery_aligning}.
	They  are all convex; non-convex shapes can also be represented by \eqref{eq:profile} for $\phi>\phic$, but they play no role in our analysis, as they do not minimize $\mathcal{F}$. 
	
	Within the class of shapes described by \eqref{eq:profile}, the aspect ratio $\delta$ of $\region$ can be expressed as an explicit function of the parameters $(\phi,\mu)$,
	\begin{equation}
		\label{eq:aspect_ratio}
		\delta(\phi,\mu):=\frac{\mu}{R(\phi,\mu;0)}=\frac{\mu^2\left(\frac83\cos\phi+\pi\sin\phi\right)}{\pi(\cos\phi+\sin\phi)}.
	\end{equation}
	The major advantage of using \eqref{eq:profile} to represent the admissible shapes of $\region$ is that the functional $\mathcal{F}$ in \eqref{eq:F_energy_functional_aligning} reduces to a function $F(\alpha,\beta;\phi,\mu)$ defined on $\conf$ for any given value of the pair $(\alpha,\beta)$. We only need to minimize $F$ over $\conf$ to obtain an approximate minimizer $\region$ of $\free$, a task that can be  accomplished numerically with fairly standard methods.
	
	\section{Geometric Method}\label{sec:method}
	We wish to apply the theory outlined in Sec.~\ref{sec:problem} to interpret 
	the experiment performed in \cite{yi:orientation}.
	There, two aligning substrates consisting in replicas of $250\,\mathrm{nm}$-deep linear channels equally spaced at $250\,\mathrm{nm}$ were overlaid parallel to one another, separated by a gap of $7\,\mu \mathrm{m}$; the cell they delimited was filled with  a DSCG solution prepared at concentration $c=11\, \mathrm{wt}\%$   by dissolving the chromonic material  in deionized water.
	At temperature $T=25^\circ \mathrm{C}$, elongated b\^atonnets were observed with a dipolar director field $\n$ on their boundaries and their long axes  aligned to the channels.
	
\begin{figure}[h]
		\centering
		\includegraphics[width=.5\linewidth]{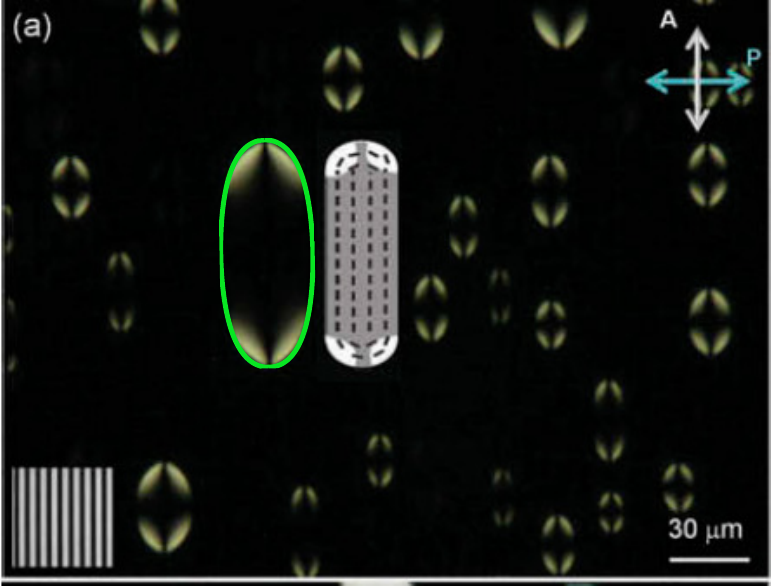}
		\caption{(Color online) Experimental picture, borrowed from Fig.~12a of \cite{yi:orientation}, representing the top view of a cell of thickness $h=7\,\mu\mathrm{m}$ filled with DSCG prepared at concentration $c=11\,\mathrm{wt}\%$ and observed between crossed polarizers (marked on the top right corner) in the coexisting biphasic regime (where the concentration of the nematic phase is estimated to be $c\approx 13.5\,\mathrm{wt}\%$). The white stripes on the bottom left corner designate the orientation of the aligning channels on both bounding substrates. The droplet  outlined in green corresponds to the minimizer $\phi_0\doteq1.77$ and $\mu_0\doteq1.51$ of the free energy $F(\alpha,\beta;\phi,\mu)$ for $\alpha=110$ and $\beta=5.5\times10^{-2}$, the latter identified so as to fit  the experimental value $\delta\approx2.4$ of the droplet's aspect ratio, as  shown in Fig.~\ref{fig:aspect_ratio_aligning}. The polar distance is $2R_0=2\mu\Req\approx96\,\mu\mathrm{m}$ (for $\Req\approx32\,\mu\mathrm{m}$). The green outline, obtained from \eqref{eq:profile}, is superimposed to the experimental image.   The whitish cartoon is an illustration (proposed in \cite{yi:orientation}) of  the observed director field; it is not meant to represent closely the droplet's shape.}
		\label{fig:Clark_experiments}
	\end{figure}

	In particular, we shall focus on the droplet shown in Fig.~12a of \cite{yi:orientation} (and highlighted in Fig.~\ref{fig:Clark_experiments}).
	Its area $A_0$ and aspect ratio $\delta$ are estimated to be
	\begin{equation}
		\label{eq:A0_delta_aligning}
		A_0\approx3217\,\mu \mathrm{m}^2, \quad\text{and}\quad \delta\approx2.4,
	\end{equation}
	the former corresponding to $\Req\approx32\,\mu\mathrm{m}$. \nigh{We read off from the phase diagram for DSCG in Fig.~2a of \cite{zhang:influence} that at $T=25^\circ \mathrm{C}$ the concentration of the coexisting nematic phase is approximately $13.5\,\mathrm{wt}\%$, larger than the concentration of the preparation (as expected). Using  the curves that in \cite{zhou:elasticity_2014} represent the temperature dependence of the elastic constants of the nematic phase of DSCG at  $c = 14\, \mathrm{wt}\%$,  we readily find that at $T=25^\circ\mathrm{C}$ 
		\begin{equation}
			\label{eq:constants_Clark}
			K_{11}\approx3\,\mathrm{pN}, \quad\text{and}\quad k_{3}\approx4.5.
		\end{equation}
		As for the isotropic surface tension $\gamma$, we take the estimate $\gamma\approx 10\,\mu\mathrm{J/m^2}$ suggested by our previous study \cite{paparini:shape}). We then obtain from \eqref{eq:constants_Clark} and \eqref{eq:alpha} that $\alpha\approx110$.}\footnote{We are aware that such an estimate for $\alpha$ may be affected by the value chosen for $\gamma$, that applies to a DSCG solution in conditions different than the ones occurring in \cite{yi:orientation}. Unfortunately, we lack better data.}
	
	Some estimates of the anchoring strength $\sigma_0$ for chromonics in contact with different substrates are already known: they range from $\sigma_0\sim10^{-1}\,\mu\mathrm{J/m^2}$, for both scratched glasses \cite{mcguire:orthogonal} and rubbed polyimide surfaces \cite{collings:anchoring}, to $\sigma_0\sim10^2\,\mu\mathrm{J/m^2}$, for surfaces lithographed by secondary sputtering \cite{kim:macroscopic}.\footnote{For thermotropic liquid crystals, the  strength of planar anchoring ranges from about $1\,\mu\mathrm{J/m^2}$ to one or two orders of magnitude higher, as shown, for example, in Table~3.1 of \cite{blinov:electrooptic}. On the strongest side is the measurement of \cite{faetti:strong}, based on an improved reflectometric method introduced in \cite{faetti:improved}; for 5CB, it was found that $\sigma_0\sim10^2\mathrm{J/m^2}$.}
	In the experiment under consideration, the aligning surfaces were lithographed, and so we expect $\sigma_0$ to be in the range  $10\text{-}10^2\,\mu\mathrm{J/m^2}$. By using $K_{11}\sim1\,\mathrm{pN}$ from \eqref{eq:constants_Clark} and $\gamma\sim10\,\mu\mathrm{J/m^2}$ from \cite{paparini:shape},  for $h\sim10\,\mu\mathrm{m}$, we estimate from \eqref{eq:beta} that $\beta\sim10^{-2}$.
	
\begin{figure}[h]
		\centering
		\includegraphics[width=0.29\linewidth]{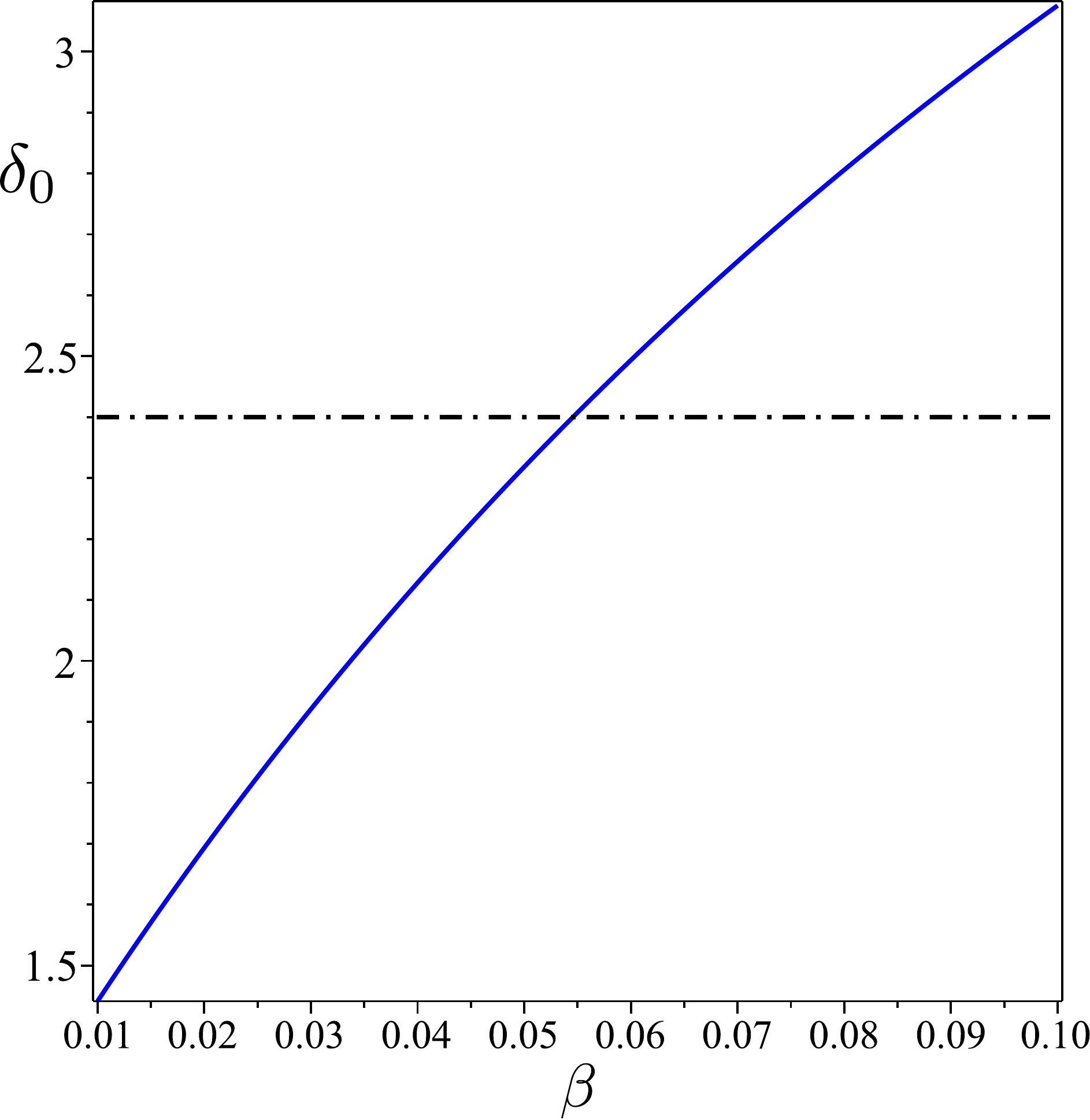}
		\caption{Graph of the aspect ratio $\delta_0=\delta(\phi_0,\mu_0)$ according to \eqref{eq:aspect_ratio} for the   minimizer $(\phi_0,\mu_0)$ of the free energy $F(\alpha,\beta;\phi,\mu)$ for $\alpha=110$, $k_3=4.5$, and $0.01\leqq\beta\leqq0.1$.}
		\label{fig:aspect_ratio_aligning}
	\end{figure}

	We thus seek numerically the minimum of $F(\alpha,\beta;\phi,\mu)$ in the pair $(\phi,\mu)$ for $\alpha=110$ and $0.01\leqq\beta\leqq0.1$; for every value of $\beta$ in this interval, we compute the theoretical value of the aspect ratio $\delta$ according to \eqref{eq:aspect_ratio}, obtaining the graph shown in Fig.~\ref{fig:aspect_ratio_aligning}.
	\nigh{The experimental value of $\delta$ is met for 
		\begin{equation}
			\label{eq:beta_Clark}
			\beta \doteq 5.5\times10^{-2}
		\end{equation}
		and the corresponding coordinates of the energy minimizer in $\conf$ are $\phi_0\doteq1.77$ and $\mu_0\doteq1.51$.} The predicted equilibrium shape is a b\^atonnet, which is  to be compared with the shape of the droplet  experimentally observed  in Fig.~\ref{fig:Clark_experiments}. Theory and experiment seem to be in good agreement. By combining \eqref{eq:beta_Clark} and \eqref{eq:beta}, we arrive at  the following estimate of the anchoring strength,
	\begin{equation}
		\label{eq:sigma_Clark}
		\nigh{\sigma_0\approx26\,\mu\mathrm{J/m^2}},
	\end{equation}
	which turns out to have the same order of magnitude as $\gamma$ and intermediate between values  measured with other methods  for the same material.
	
	\subsection{Tangential Anchoring Breaking}\label{sec:alpha_safeguard}
	As in our previous work \cite{paparini:shape}, the theory presented here is based on the assumption that the director configuration at the boundary of the drop remains bipolar for all admissible values of the area $A_0$.
	
	Although the validity of  this constraint is confirmed experimentally, we are aware that it cannot hold for all values of $A_0$, as the tangential anchoring  of $\n$ is bound to be broken both for $A_0$ sufficiently small and for $A_0$ sufficiently large. Indeed, for given $h$,  the elastic, interfacial, and anchoring energies scale like $\Req^0$, $\Req^1$, and $\Req^2$, respectively (see also \cite{virga:drops}, for a similar reasoning). Thus, for $\alpha\ll1$, the elastic energy dominates, promoting a uniform orientation of $\n$, preferentially along the aligning channels. On the other hand, for $\alpha\gg1$, the anchoring energy dominates, promoting the same uniform alignment of $\n$. In both limiting cases, the tangential anchoring at the isotropic interface is broken. 
	
\begin{figure}[h]
		\centering
		\begin{subfigure}[b]{0.30\linewidth}
			\centering
			\includegraphics[width=.95\linewidth]{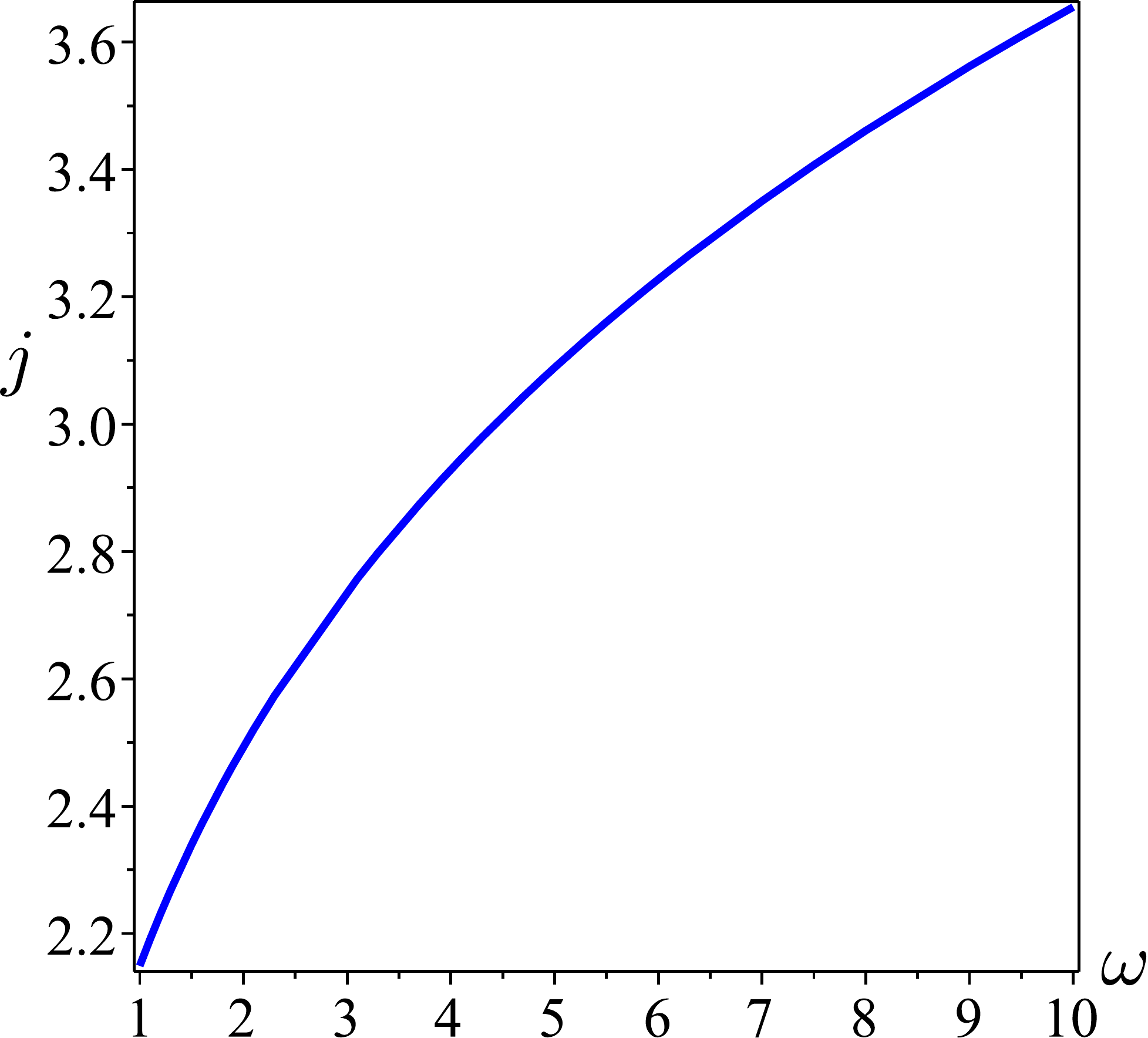}
			\vskip10pt
			\caption{Graph  of the function $j(\omega)$ (from Fig.\,13b of \cite{paparini:shape}).}
			\label{fig:j_graph}
		\end{subfigure}
		\qquad
		\begin{subfigure}[b]{0.32\linewidth}
			\includegraphics[width=.9\linewidth]{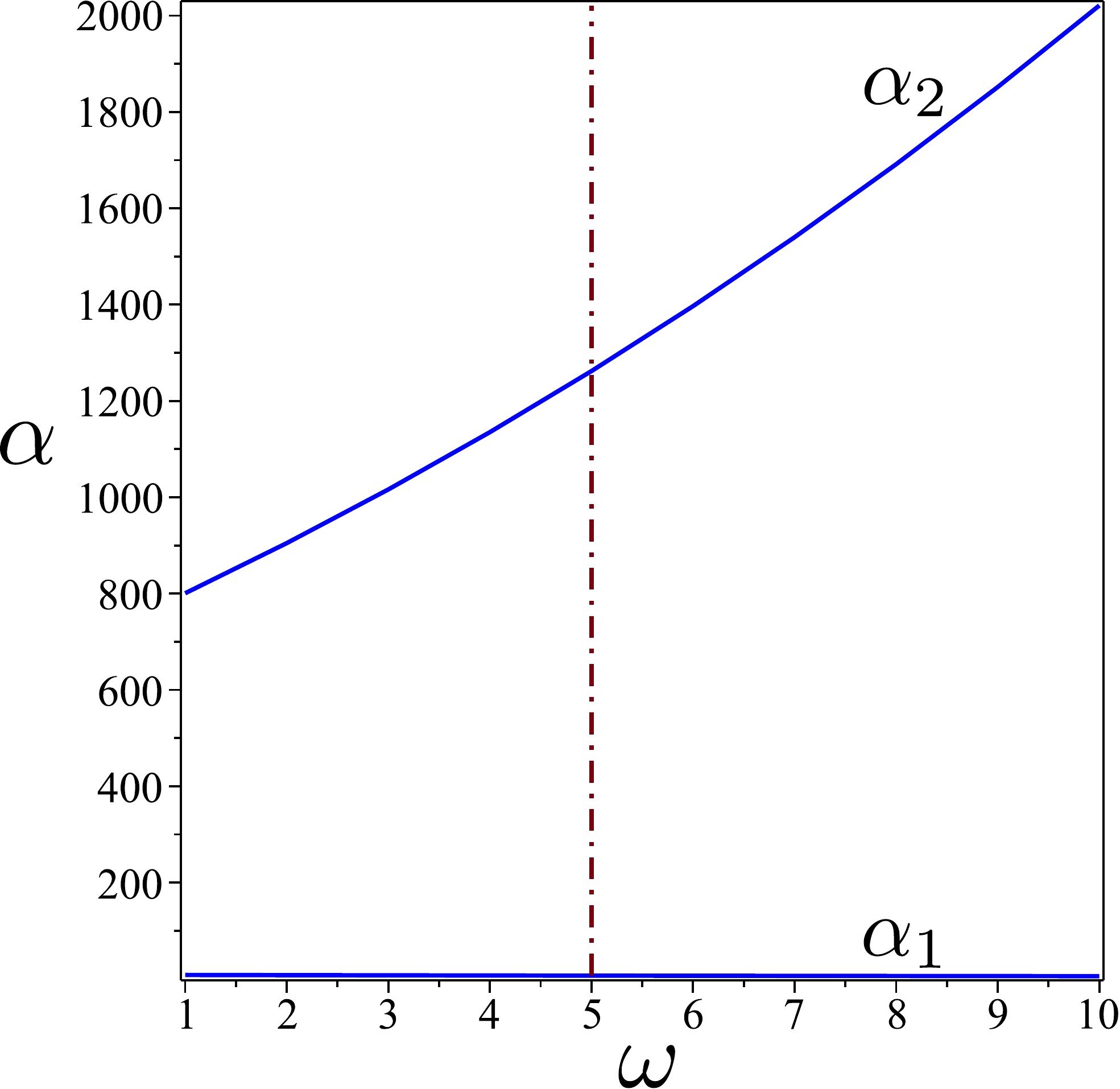}
			\caption{Farthest apart roots of the polynomial $P_\lambda$ in \eqref{eq:function_alpha_safeguard} plotted against $\omega$.}
			\label{fig:safeguard_alignment}
		\end{subfigure}
		\caption{Roots of polynomial $P_\lambda(\alpha)$ in \eqref{eq:function_alpha_safeguard} as $\lambda$ ranges in the interval $(0,1]$,  for $\varepsilon=10^{-3}$ and $\beta$ and $k_3$ as in \eqref{eq:beta_Clark} and \eqref{eq:constants_Clark}, respectively.}
	\end{figure}

	Direct energy comparisons performed with the method illustrated in Appendix~B of \cite{paparini:shape} (based on constructions by Wulff~\cite{wulff:frage} and Williams~\cite{williams:transitions}) allowed us to estimate an interval  $\alpha_1\leqq\alpha\leqq\alpha_2$, within which we can be confident that the tangential anchoring hypothesized here on the isotropic interface is \emph{not} broken. The end-points of such a \emph{safeguard} interval for $\alpha$ were identified as the farthest apart roots of the following polynomial in $\alpha$ as $\lambda$ ranges in the interval $0<\lambda\leqq1$,
	\begin{equation}
		\label{eq:function_alpha_safeguard}
		P_\lambda(\alpha):=
		-\frac13\beta\lambda^2\alpha^2+\left[4j(\omega)-4\lambda\left(\frac{\pi}{2}-\varepsilon\right)-\frac{\pi}{\lambda}\left(1-\lambda^2\right)-2\omega\lambda\pi\varepsilon\right]\alpha
		-4\left[\frac{\pi}{4}\left(k_3-1-k_3\ln2-\ln\varepsilon\right)+\varepsilon\right],
	\end{equation}
	where $j$ is a monotone function of the dimensionless anisotropic strength $\omega$ in \eqref{eq:Rapini_Papoular_formula} (see Fig.~\ref{fig:j_graph}). Figure~\ref{fig:safeguard_alignment}
	shows how both $\alpha_1$ and $\alpha_2$ depend on $\omega$ (the latter more dramatically than the former) for $\varepsilon=10^{-3}$ and $\beta$ and $k_3$ as in \eqref{eq:beta_Clark} and \eqref{eq:constants_Clark}, respectively. Letting $\omega=5$, which is a choice supported by some evidence \cite{puech:nematic,kim:morphogenesis}, we obtain that $\alpha_1\doteq7$ and $\alpha_2\doteq1262$, showing that the case we have studied ($\alpha\approx110$) falls well inside the range of validity of our model.\footnote{Reverting the safeguard interval for $\alpha$ into one for the equivalent radius $\Req$, we arrive at $0.1\,\mu\mathrm{m}\lessapprox\Req\lessapprox0.4\,\mathrm{mm}$, confirming again that in the case of interest ($\Req\approx32\,\mu\mathrm{m}$) our theory is perfectly legitimate.}
	
	In the following section, we shall explore the equilibrium shapes of $\region$ for values of $\alpha$ both smaller and larger than the one corresponding to  the experimental shape outlined in Fig.~\ref{fig:Clark_experiments}, but still within the safeguard interval identified above. We shall see that for sufficiently small droplets our theory also predicts \emph{shape bistability} in the present setting  of aligning substrates, as it did in \cite{paparini:shape} for planar degenerate ones.
	
	\section{Shape Bistability}\label{sec:bistability}
	We extended the analysis of the minimizers of the reduced free energy function $F(\alpha,\beta;\phi,\mu)$ by allowing $\alpha$ to cover the whole safeguard interval corresponding to the values of $k_3$ and $\beta$ in \eqref{eq:constants_Clark} and \eqref{eq:beta_Clark}, respectively. Our aim was to see whether our theory would also predict droplets' equilibrium shapes qualitatively different than those observed in \cite{yi:orientation} (representative examples of which are reported in Fig.~\ref{fig:Clark_experiments}). We found out that it \emph{does}, in a range of sufficiently small values of the droplets' area; they do not seem to have been observed, at least in \cite{yi:orientation}. 
	
\begin{figure}[h]
		\centering
		\begin{subfigure}[b]{0.31\linewidth}
			\centering
			\includegraphics[width=\linewidth]{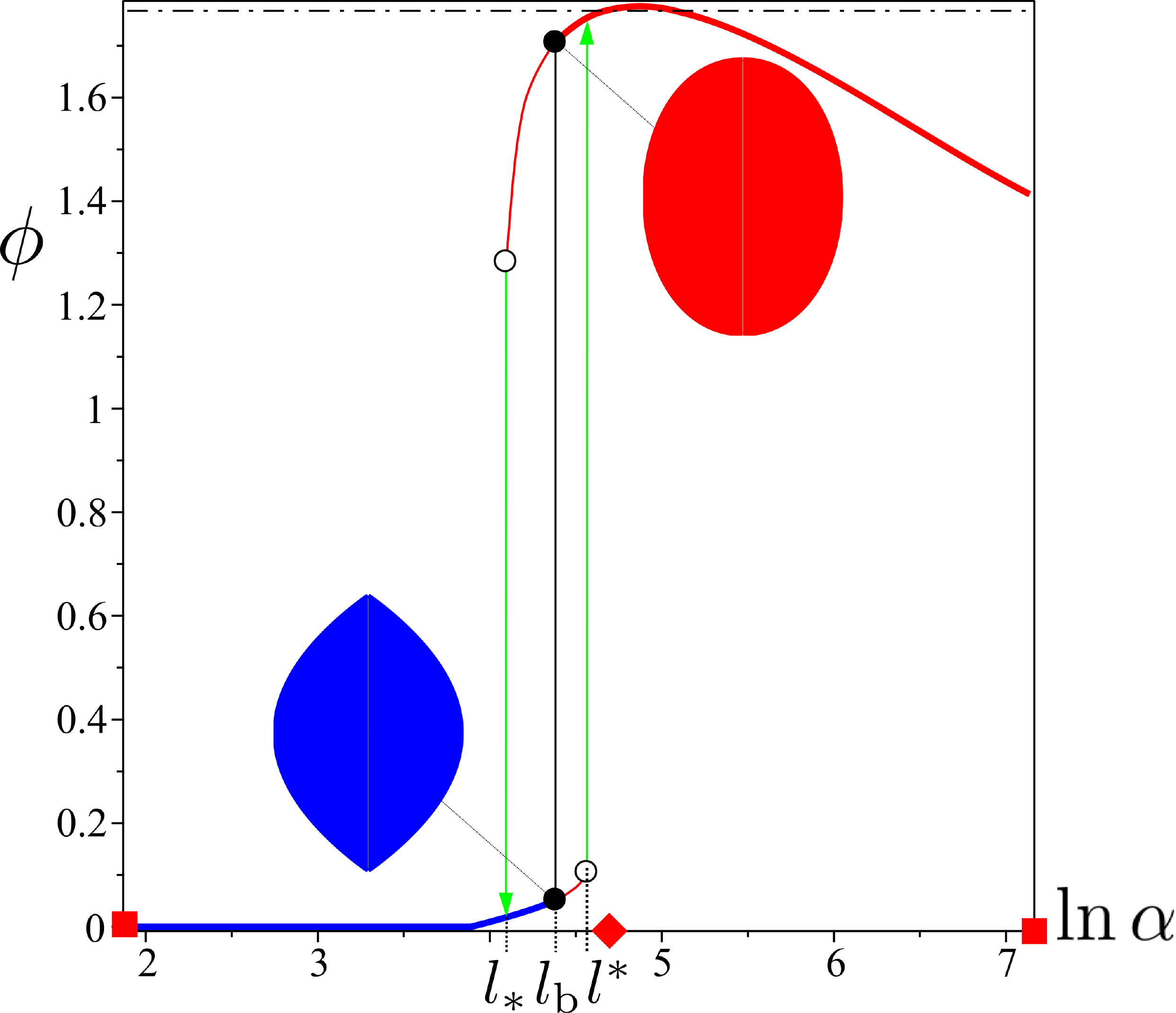}
			\caption{Minimizers' $\phi$  trajectory}
			\label{fig:phi_minima_alignment}
		\end{subfigure}
		\begin{subfigure}[b]{0.31\linewidth}
			\centering
			\includegraphics[width=\linewidth]{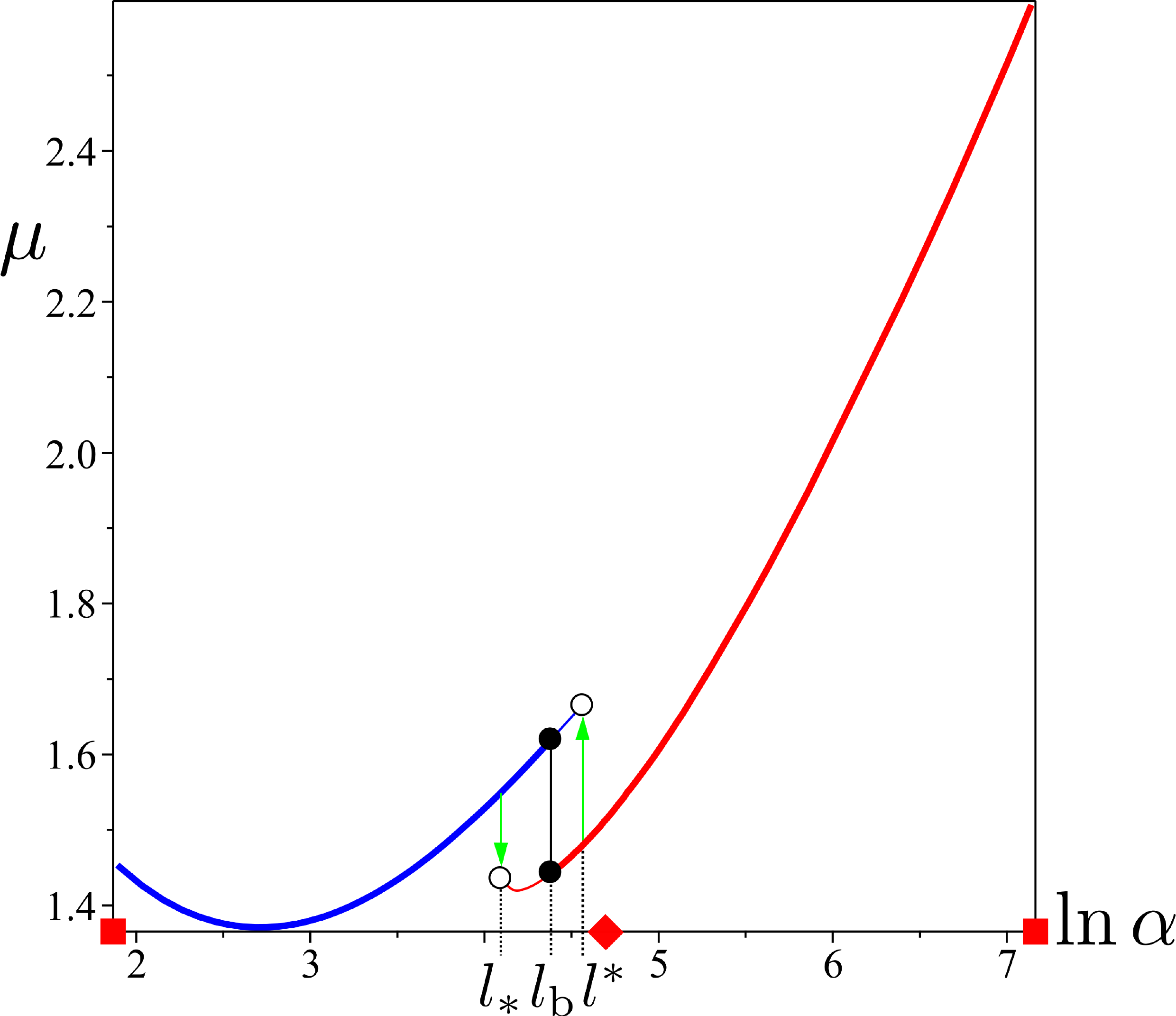}
			\caption{Minimizers' $\mu$  trajectory}
			\label{fig:mu_minima_alignment}
		\end{subfigure}
		\begin{subfigure}[b]{0.34\linewidth}
			\centering
			\includegraphics[width=.94\linewidth]{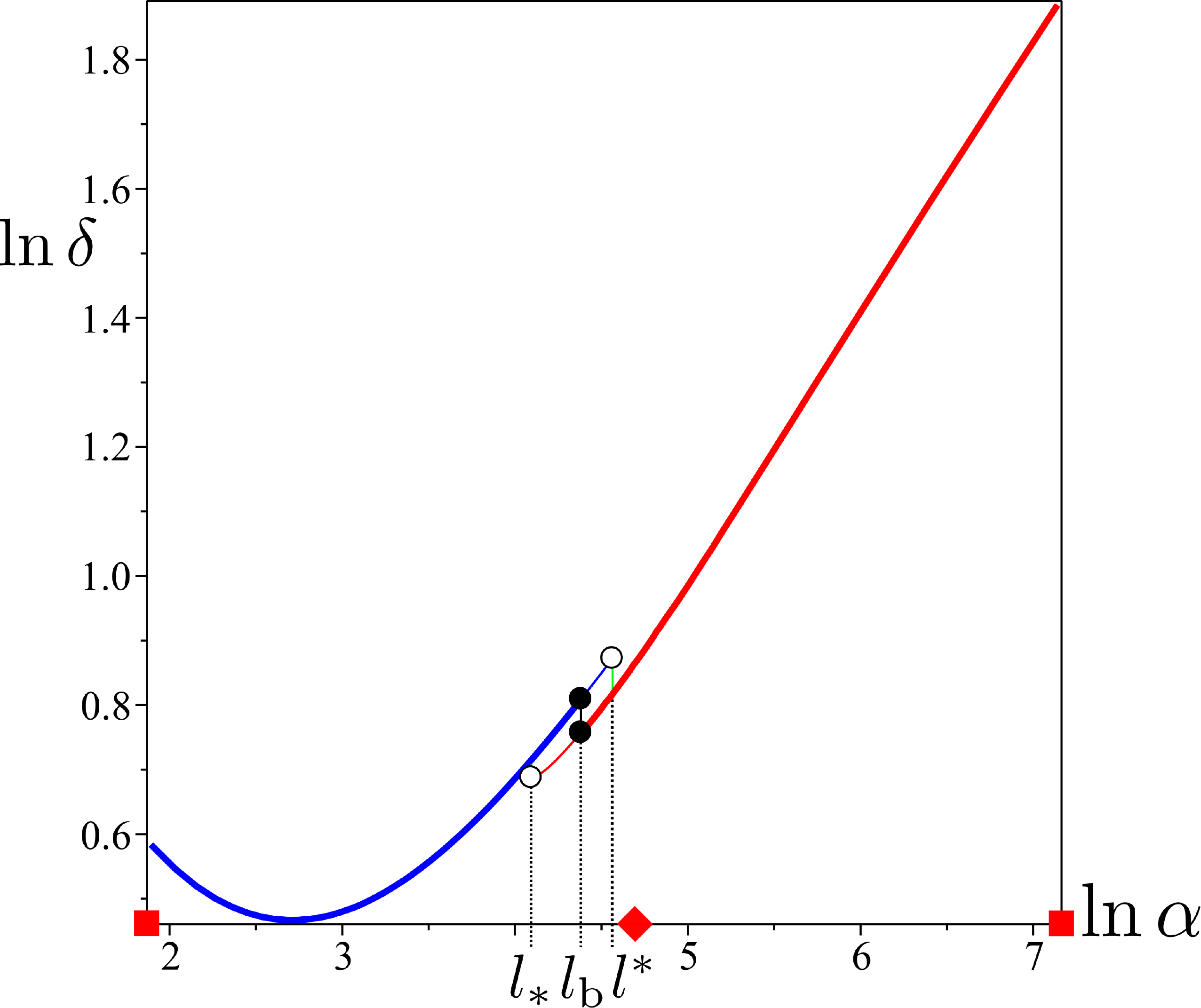}
			\caption{Droplet's aspect ratio}
			\label{fig:aspect_ratio_bifurcation}
		\end{subfigure}
		\caption{(Color online) Bifurcation diagrams with hysteresis of the energy minimizers drawn for $k_3=4.5$ and $\beta=5.5\times10^{-2}$; thick lines represent global minimizers (stable branches) of $F(\alpha,\beta;\phi,\mu)$ in the pair $(\phi,\mu)$, while thin lines represent local minimizers (metastable branches). Open circles mark  the equilibrium shapes that delimit the interval of coexistence of tactoidal and discoidal    shapes, while green lines bound the corresponding shape hysteresis. Black dots identify the two perfectly bistable minimizers.The end-points $\alpha_1$ and $\alpha_2$ of the safeguard interval for $\alpha$ are marked by (red) squares, while the value $\alpha\approx110$ corresponding to the shape outlined in Fig.~\ref{fig:Clark_experiments} is marked by a (red) diamond. In panel (a), a  red dot signals the value of $\alpha$ where the trajectory of minimizers leaves the $\phi=0$ axis and equilibrium tactoids cease to be genuine. The conventional border of b\^atonnets is placed at $\phi\doteq1.77$, while the tactoidal territory is delimited by the conventional border at $\phi\doteq0.20$. Here, $\alphadown\approx60$,  $\alphab\approx80$, and  $\alphaup\approx96$; the following abbreviations are used on the horizontal axis: $l_*:=\ln\alphadown\doteq4.1$, $l_\mathrm{b}:=\ln\alphab\doteq4.4$, and $l^*:=\ln\alphaup\doteq4.6$.}
		\label{fig:phimu_alpha_shape}
	\end{figure}

	In brief, we found that, for $\alpha$ in the interval $\alphadown\leqq\alpha\leqq\alphaup$, \emph{two} equilibrium shapes compete for the global energy minimizer, a tactoid and a discoid, coexisting as local minimizers and exchanging their role as global minimizer at a critical value, $\alpha=\alphab$, of perfect bistability. Details of our analysis are illustrated in the bifurcation diagrams with hysteresis shown in Fig.~\ref{fig:phimu_alpha_shape}.
	
	For $\alpha<\alphadown$, only the (blue) tactoidal branch exists and is globally stable. As soon as $\alpha$ exceeds $\alphadown$ the (red) discoidal branch comes into life as a metastable equilibrium, and takes over the tactoidal branch as energy minimizer  at $\alpha=\alphab$. Two black dots mark the exchange of stability occurring in the system; close to them in Fig.~\ref{fig:phi_minima_alignment} are placed  the corresponding equilibrium shapes depicted in the same color as the equilibrium branch they belong to. For $\alpha>\alphab$, the metastable tactoidal branch ceases altogether to exist  at $\alpha=\alphaup$ and gives way to the discoidal one as unique equilibrium branch.  In an interval, the $\phi$ trajectory of energy minimizers in Fig.~\ref{fig:phi_minima_alignment} traverses the b\^atonnet border, while staying otherwise in the discoidal territory for $\alpha>\alphaup$. For $\alpha<\alphadown$, instead, the $\phi$ trajectory stays consistently within the tactoidal territory.
	
	Two features deserve notice. First, discoids are energy minimizers, but not for  all values of $\alpha$: upon increasing the droplet's area, the equilibrium shape undergoes two smooth transitions, from a discoid to a b\^atonnet and back again to a discoid.
	Second, as shown by both  Figs.~\ref{fig:mu_minima_alignment} and \ref{fig:aspect_ratio_bifurcation}, upon increasing the droplet's area in the whole admissible domain, the equilibrium shape  is first  thickened   and then thinned, suffering a transient setback at the transition.
	
	The  transition  values of $\alpha$ shown in Fig.~\ref{fig:phimu_alpha_shape} are $\alphadown\approx60$,  $\alphab\approx 80$, and  $\alphaup\approx96$,  while the re-entrant b\^{a}tonnet interval in Fig.~\ref{fig:phi_minima_alignment} is   $106\leqq\alpha\leqq160$, where in particular falls the value $\alpha\approx110$ corresponding to the droplet outlined in Fig.~\ref{fig:Clark_experiments}. In physical units,  according to the model proposed here, one would then expect coexistence of tactoids and discoids for $18\,\mu\mathrm{m}\lessapprox \Req\lessapprox 29\,\mu\mathrm{m}$, a regime of small droplets for which no data are available in \cite{yi:orientation}. For $\Req\gtrapprox29\,\mu\mathrm{m}$, only non-tactoidal shapes should be observed; they are discoids, except for $32\,\mu\mathrm{m}\lessapprox\Req\lessapprox48\,\mu\mathrm{m}$, where they are b\^atonnets.

	\section{Conclusion}\label{sec:conclusion}
	We proposed a method to determine the planar anchoring strength $\sigma_0$ of a chromonic liquid crystal on a rigid substrate; its distinctive feature  is  geometric,  as it is based on the observation and fitting of the stable equilibrium shapes of droplets in the nematic phase coexisting in a cell with the isotropic phase. Prior knowledge of the surface tension $\gamma$ of the nematic phase at the isotropic interface is presumed, which can be gained by use of cells with substrates enforcing planar degenerate anchoring \cite{paparini:shape}. 
	
	Our study was motivated by the experiment described in \cite{yi:orientation}, where nematic chromonic droplets formed in a thin cell enclosed within parallel planar substrates orienting the  director $\n$ in one and the same direction. To illustrate our method, we applied it to one of the DSCG droplets shown in \cite{yi:orientation} and extracted  an estimate for $\sigma_0$ from its shape.  Although this figure for $\sigma_0$ is similar to those obtained with other methods, by no means can our estimate be regarded as a \emph{measure} of $ \sigma_0$, as it lacks the appropriate statistics and, what is perhaps more important, an independent determination of $\gamma$.
	
	\nigh{Opting for importing the value of $\gamma$ from other sources, we determined $\beta$ (and thus $\sigma_0$) by fitting the aspect ratio of a selected, representative droplet, judging then the agreement between experiment and  theory from a qualitative comparison between observed and predicted shapes. This suffices to provide a 
		proof of principle that the proposed method is indeed viable. It can be improved: using $\alpha$ and $\widetilde{\beta}=2\alpha^2\beta$ as independent  parameters,  one could determine \emph{both} $\gamma$ and $\sigma_0$ by fitting an observed shape with the family described in \eqref{eq:profile} by the parameters $(\phi,\mu)$ (possibly through a shape recognition algorithm). The optimal $(\alpha,\widetilde{\beta})$ would then be determined by minimizing the free energy  $F(\alpha,\widetilde{\beta};\phi_0,\mu_0)$ associated with the best shape represented by $(\phi_0,\mu_0)$. Doing this for several droplets (each with its own $\Req$) would result in several measures of both $\gamma$ and $\sigma_0$.}
	
	One method successfully employed so far to measure $\sigma_0$ for chromonics uses twist cells with plates promoting planar easy axes at right angles to one another. Measuring the total \emph{twist angle} $\Omega$ across the cell (and how it differs from $90^\circ$) determines $\sigma_0$, once the twist constant $K_{22}$ is known  \cite{mcginn:planar}. This method relies on the theory (put forward  by McIntyre~\cite{mcintyre:light,mcintyre:transmission}) relating (in closed form) $\Omega$ to the maximum and minimum transmitted intensity of light with normal incidence propagating (between crossed polarizers) through the cell. Although this theory has a wider range of validity than Mauguin's \emph{adiabatic limit} \cite{mauguin:representation} (see also \cite[p.\,268]{degennes:physics}), it makes approximations too. It might thus be valuable to have an alternative, independent method to rely upon.
	
	Our theory was developed having especially chromonic nematics in mind, but nothing prevents one from applying it to thermotropic nematics as well, as in the setting envisioned here no macroscopic differences arise between these materials.
	
	The droplets' shapes observed in \cite{yi:orientation}  mainly resembled elongated rods with rounded ends, which we called b\^atonnets. Theory also predicted other stable equilibrium shapes: either slender and round, which we called discoids, or slender and pointed, which we called tactoids. Moreover, in the range of \emph{small} droplets, we found a regime of bistability, where discoids and tactoids coexist as energy minimizers, taking turns in being alternatively stable or metastable. A similar bistability was also predicted for two-dimensional droplets between parallel plates enforcing planar degenerate anchoring \cite{paparini:shape}, but not for fully three-dimensional droplets \cite{paparini:nematic}. Although these two coincidences \emph{cannot} be a proof,\footnote{Common wisdom has it that Agatha Christie once said that one coincidence is a coincidence, two coincidences are a clue, three coincidences a proof. We could not locate this precise quote in her writings, but in \emph{The ABC Murders} Mr.~Poirot comes close when he says : ``It is the same motif three times repeated. That cannot be coincidence.''} we are inclined to think that bistability of shape might be a \emph{two-dimensionality} signature.

	\begin{acknowledgements}
		We are grateful to the Reviewers for the constructive criticism that has improved our work.
	\end{acknowledgements}	
	
	\appendix 
	
	\section{Further Mathematical Details}\label{sec:retracted_field}
	This Appendix contains ancillary calculations used to arrive at the dimensionless form of $\free$ in \eqref{eq:F_energy_functional} for the retracted bipolar vector field $\n$ described in Sec. \ref{sec:admissible_shapes}. A fuller account can be found in \cite{paparini:shape}. The boundary curve $R=R(y)$ is retracted inside $\region$ as the curve
	\begin{equation}
		\label{eq:curve_p_t}
		\bm{p}_t(y):=g(t)R(y)\e_x+y\e_y, \quad-R_0\leqq y\leqq R_0,
	\end{equation}
	where $t\in[0,1]$ and $g$ is any strictly increasing function on $[0,1]$ of class $\mathcal{C}^1$ such that $g(0)=0$ and $g(1)=1$. In this two-dimensional setting, $\n$ is defined as the unit vector field tangent to the lines represented by \eqref{eq:curve_p_t} at fixed $t$; by differentiating $\bm{p}_t$ in \eqref{eq:curve_p_t} with respect to  $y$, keeping $t$ fixed, we easily obtain that
	\begin{eqnarray}
		\label{eq:n}
		\n=\frac{gR'\e_x+\e_y}{\sqrt{1+(gR')^2}},
	\end{eqnarray}
	where a prime denotes differentiation. 
	The element of area $\dd A$ of $\region$ is  given by
	\begin{equation}
		\label{eq:dA}
		\dd A=\dd t\dd y g'R\sqrt{1+(gR')^2}\e_x\times\n\cdot\e_z =g'R\dd t\dd y.
	\end{equation}
	
	By rescaling all lengths to the radius $\Req$ of the disc of area $A_0$, we obtain the following dimensionless form of the total anchoring energy in \eqref{eq:free_energy_functional_aligning},
	\begin{equation}
		\label{eq:F_energy_functional_t} 
		\mathcal{F}_\mathrm{a}[\mu;R]:=-\frac{\sigma_0}{K_{11}h}\int_{\region}(\n\cdot\e_y)^2\dd A=-2\frac{\sigma_0\Req^2}{K_{11}h}\int_{-\mu}^{\mu} \dd y\int_0^1 \frac{g'R}{1+(gR')^2}\dd t,
	\end{equation}
	where $\mu$ is defined in \eqref{eq:mu_definition}. By use of \eqref{eq:alpha} and \eqref{eq:beta}, we readily give \eqref{eq:F_energy_functional_t} the form in \eqref{eq:free_sub_aligning_scaled}, which is independent of the specific choice of $g$, provided it is monotonic and obeys the prescribed boundary conditions.
	\nigh{
		\section{Boojum's defect core}\label{sec:core_boojum}
		Here we estimate the energy stored in the defect core of a boojum on the boundary of a two-dimensional droplet. To this end, we apply a simplified version of Ericksen's model \cite{ericksen:liquid}. We write the total free energy in the form
		\begin{equation}\label{eq:core_energy}
			\free_\mathrm{c}[S,\n]:=h\int_{\core}\left\{\frac{K}{2}\left(k|\nabla S|^2+S^2|\nabla\n|^2\right)+\psi(S)\right\}\dd A,
		\end{equation}
		where $S$ denotes the scalar order parameter and $\n$ is the director field, $\psi$ is the \emph{condensation} potential$, K$ in an average elastic constant and $0<k<1$ is a dimensionless parameter.  The domain of integration $\core$ is a circular sector of radius $r_\mathrm{c}=\varepsilon\Req$ with (inner) \emph{cusp angle} $\tau$, located at each pole of the region $\region$ shown in Fig.~\ref{fig:shape_clark}: $\tau=\pi$ when $\partial\region$ is smooth, that is, for non-genuine tactoids, discoids, and b\^atonnets (see Fig.~\ref{fig:gallery_aligning}), whereas $\tau<\pi$ for \emph{genuine} tactoids. We write the condensation potential $\psi$ in the standard form,
		\begin{equation}
			\label{eq:condensation_potential}
			\psi(S)=\frac12aS^2-\frac13bS^3+\frac14cS^4,
		\end{equation}
		where $a$ depends on temperature $T$ as $a=A(T-T^\ast)$, $A$ is a positive constant and $T^\ast$ is the \emph{supercooling} temperature, while both $b$ and $c$ are positive constant independent of temperature.
		
		The representation for $\free_\mathrm{c}$ in \eqref{eq:core_energy} is valid under the assumption that $S$ and $\n$ are independent of the coordinate $z$ across the cell confining the droplet. For our estimate, we  further take $\n=\e_r$, where $\e_r$ is the unit vector field in the plane $(x,y)$ emanating from the pole of $\region$, and we assume that $S=S(r)$, where $r$ is the radial coordinate, subject to the boundary condition
		\begin{equation}
			\label{eq:S_boundary condition}
			S(r_\mathrm{c})=S_0.
		\end{equation}
		Here, for $b^2>4ac$, 
		\begin{equation}
			\label{eq:S_0}
			S_0:=\frac{b+\sqrt{b^2-4ac}}{2c}
		\end{equation}
		is the absolute minimizer of $\psi$.
		
		Under these assumptions, by scaling lengths to $r_\mathrm{c}$, we give $\free_\mathrm{c}$ the following dimensionless form
		\begin{equation}
			\label{eq:core_energy_dimensionless}
			\mathcal{F}_\mathrm{c}[S]:=\frac{1}{Kh}\free_\mathrm{c}[S,\e_r]=\tau\int_0^1\left\{\frac12kS'^2+\frac12\frac{S^2}{\rho^2}+\xi\left(\frac12S^2-\frac13b_0S^3+\frac14c_0S^4\right)\right\}\rho\dd\rho,
		\end{equation}
		where
		\begin{equation}
			\rho:=\frac{r}{r_\mathrm{c}},\quad\xi:=\frac{ar_\mathrm{c}^2}{K},\quad b_0:=\frac{b}{a},\quad c_0:=\frac{c}{a},
		\end{equation}
		and a prime denotes differentiation with respect to $\rho$. In \eqref{eq:core_energy_dimensionless}, $\xi$ weights the condensation energy against the elastic energy. To estimate this dimensionless parameter, we resort to the classical data of \cite{poggi:free} (see also \cite[p.\,130]{kleman:soft}) and for $T-T^\ast\sim1\,\mathrm{K}$, $r_\mathrm{c}\sim10\,\mathrm{nm}$, and $K\sim1\,\mathrm{pN}$, we find that $\xi\sim1$.\footnote{More precisely, for MBBA, we learn from \cite{poggi:free} that $\frac12A\approx5\times10^4\,\mathrm{Jm^{-3}K^{-1}}$, $\frac13b\approx2\times10^5\,\mathrm{Jm^{-3}}$, $\frac14c\approx3\times10^5\,\mathrm{Jm^{-3}}$. Since $S\sim10^{-1}$, for $T-T^\ast>1\,\mathrm{K}$, the order of magnitude of $\psi(S)$ is that of $aS^2$.} Thus, elastic and condensation components of the free energy in \eqref{eq:core_energy_dimensionless} have the same order of magnitude. Since $\psi(S)\geqq\psi(S_0)$ for $0\leqq S\leqq S_0$, $\mathcal{F}_\mathrm{c}$ satisfies the inequality
		\begin{equation}
			\label{eq:core_energy_inequality}
			\mathcal{F}_\mathrm{c}[S]\geqq\frac12\tau\int_0^1\left(kS'^2+\frac{S^2}{\rho^2}\right)\rho\dd\rho+\frac12\tau\xi S_0^2\left(\frac12-\frac13b_0S_0+\frac14c_0S_0^2\right).
		\end{equation}
		
		An estimate for the the minimum of $\mathcal{F}_\mathrm{c}$ can be obtained by minimizing the integral on the right-hand side of \eqref{eq:core_energy_inequality} subject to the boundary conditions
		\begin{equation}
			\label{eq:core_boundary_conditions}
			S(1)=S_0\quad\text{and}\quad S(0)=0,
		\end{equation}
		the former stemming from \eqref{eq:S_0} and the latter being required for $\mathcal{F}_\mathrm{c}$ to be finite. The corresponding equilibrium equation is simply given by
		\begin{equation}
			\label{eq:core_equilibrium_equation}
			k(\rho S')'=\frac{S}{\rho}.
		\end{equation}
		The solution to \eqref{eq:core_equilibrium_equation} and \eqref{eq:core_boundary_conditions} is
		\begin{equation}
			\label{eq:core_equilibrium_solution}
			S=S_0\rho^{1/\sqrt{k}}.
		\end{equation}
		We thus conclude that the core energy $\mathcal{F}_\mathrm{c}$ can be estimated from the inequality
		\begin{equation}
			\label{eq:core_energy_estimate}
			\min{\mathcal{F}_\mathrm{c}[S]}\geqq\frac12\tau S_0^2\left\{\sqrt{k}+\xi+\frac12-\frac13b_0S_0+\frac14c_0S_0^2\right\},
		\end{equation}
		which substantiates our claim in Sec.~\ref{sec:admissible_shapes} to the effect that $\mathcal{F}_\mathrm{c}$  does not depend on $r_\mathrm{c}$, but only on $S_0$ and  the cusp angle $\tau$. Moreover, since, by \eqref{eq:free_sub_aligning_scaled}, $\mathcal{F}_\mathrm{a}\sim\alpha^2\beta\mu\sim10^2\text{-}10^3$ and, by \eqref{eq:core_energy_estimate}, $\mathcal{F}_\mathrm{c}\sim10^{-2}$, it is justified to neglect $\mathcal{F}_\mathrm{c}$ in the (dimensionless) total free energy $\mathcal{F}$ in \eqref{eq:F_energy_functional_aligning}.}


\begin{thebibliography}{61}%
\makeatletter
\providecommand \@ifxundefined [1]{%
 \@ifx{#1\undefined}
}%
\providecommand \@ifnum [1]{%
 \ifnum #1\expandafter \@firstoftwo
 \else \expandafter \@secondoftwo
 \fi
}%
\providecommand \@ifx [1]{%
 \ifx #1\expandafter \@firstoftwo
 \else \expandafter \@secondoftwo
 \fi
}%
\providecommand \natexlab [1]{#1}%
\providecommand \enquote  [1]{``#1''}%
\providecommand \bibnamefont  [1]{#1}%
\providecommand \bibfnamefont [1]{#1}%
\providecommand \citenamefont [1]{#1}%
\providecommand \href@noop [0]{\@secondoftwo}%
\providecommand \href [0]{\begingroup \@sanitize@url \@href}%
\providecommand \@href[1]{\@@startlink{#1}\@@href}%
\providecommand \@@href[1]{\endgroup#1\@@endlink}%
\providecommand \@sanitize@url [0]{\catcode `\\12\catcode `\$12\catcode
  `\&12\catcode `\#12\catcode `\^12\catcode `\_12\catcode `\%12\relax}%
\providecommand \@@startlink[1]{}%
\providecommand \@@endlink[0]{}%
\providecommand \url  [0]{\begingroup\@sanitize@url \@url }%
\providecommand \@url [1]{\endgroup\@href {#1}{\urlprefix }}%
\providecommand \urlprefix  [0]{URL }%
\providecommand \Eprint [0]{\href }%
\providecommand \doibase [0]{https://doi.org/}%
\providecommand \selectlanguage [0]{\@gobble}%
\providecommand \bibinfo  [0]{\@secondoftwo}%
\providecommand \bibfield  [0]{\@secondoftwo}%
\providecommand \translation [1]{[#1]}%
\providecommand \BibitemOpen [0]{}%
\providecommand \bibitemStop [0]{}%
\providecommand \bibitemNoStop [0]{.\EOS\space}%
\providecommand \EOS [0]{\spacefactor3000\relax}%
\providecommand \BibitemShut  [1]{\csname bibitem#1\endcsname}%
\let\auto@bib@innerbib\@empty
\bibitem [{\citenamefont {Lydon}(2011)}]{lydon:chromonic}%
  \BibitemOpen
  \bibfield  {author} {\bibinfo {author} {\bibfnamefont {J.}~\bibnamefont
  {Lydon}},\ }\bibfield  {title} {\bibinfo {title} {Chromonic liquid
  crystalline phases},\ }\href {https://doi.org/10.1080/02678292.2011.614720}
  {\bibfield  {journal} {\bibinfo  {journal} {Liq. Cryst.}\ }\textbf {\bibinfo
  {volume} {38}},\ \bibinfo {pages} {1663} (\bibinfo {year}
  {2011})}\BibitemShut {NoStop}%
\bibitem [{\citenamefont {Lydon}(1998{\natexlab{a}})}]{lydon:handbook}%
  \BibitemOpen
  \bibfield  {author} {\bibinfo {author} {\bibfnamefont {J.}~\bibnamefont
  {Lydon}},\ }\bibfield  {title} {\bibinfo {title} {Chromonics},\ }in\ \href
  {https://doi.org/https://doi.org/10.1002/9783527619276.ch15c} {\emph
  {\bibinfo {booktitle} {Handbook of Liquid Crystals: {L}ow Molecular Weight
  Liquid Crystals {II}}}},\ \bibinfo {editor} {edited by\ \bibinfo {editor}
  {\bibfnamefont {D.}~\bibnamefont {Demus}}, \bibinfo {editor} {\bibfnamefont
  {J.}~\bibnamefont {Goodby}}, \bibinfo {editor} {\bibfnamefont {G.~W.}\
  \bibnamefont {Gray}}, \bibinfo {editor} {\bibfnamefont {H.-W.}\ \bibnamefont
  {Spiess}},\ and\ \bibinfo {editor} {\bibfnamefont {V.}~\bibnamefont {Vill}}}\
  (\bibinfo  {publisher} {John Wiley \& Sons},\ \bibinfo {address} {Weinheim,
  Germany},\ \bibinfo {year} {1998})\ Chap.\ \bibinfo {chapter} {XVIII}, pp.\
  \bibinfo {pages} {981--1007}\BibitemShut {NoStop}%
\bibitem [{\citenamefont {Lydon}(1998{\natexlab{b}})}]{lydon:chromonic_1998}%
  \BibitemOpen
  \bibfield  {author} {\bibinfo {author} {\bibfnamefont {J.}~\bibnamefont
  {Lydon}},\ }\bibfield  {title} {\bibinfo {title} {Chromonic liquid crystal
  phases},\ }\href
  {https://doi.org/https://doi.org/10.1016/S1359-0294(98)80019-8} {\bibfield
  {journal} {\bibinfo  {journal} {Curr. Opin. Colloid Interface Sci.}\ }\textbf
  {\bibinfo {volume} {3}},\ \bibinfo {pages} {458} (\bibinfo {year}
  {1998}{\natexlab{b}})}\BibitemShut {NoStop}%
\bibitem [{\citenamefont {Lydon}(2010)}]{lydon:chromonic_2010}%
  \BibitemOpen
  \bibfield  {author} {\bibinfo {author} {\bibfnamefont {J.}~\bibnamefont
  {Lydon}},\ }\bibfield  {title} {\bibinfo {title} {Chromonic review},\ }\href
  {https://doi.org/10.1039/B926374H} {\bibfield  {journal} {\bibinfo  {journal}
  {J. Mater. Chem.}\ }\textbf {\bibinfo {volume} {20}},\ \bibinfo {pages}
  {10071} (\bibinfo {year} {2010})}\BibitemShut {NoStop}%
\bibitem [{\citenamefont {Dierking}\ and\ \citenamefont {Martins
  Figueiredo~Neto}(2020)}]{dierking:novel}%
  \BibitemOpen
  \bibfield  {author} {\bibinfo {author} {\bibfnamefont {I.}~\bibnamefont
  {Dierking}}\ and\ \bibinfo {author} {\bibfnamefont {A.}~\bibnamefont {Martins
  Figueiredo~Neto}},\ }\bibfield  {title} {\bibinfo {title} {Novel trends in
  lyotropic liquid crystals},\ }\href {https://doi.org/10.3390/cryst10070604}
  {\bibfield  {journal} {\bibinfo  {journal} {Crystals}\ }\textbf {\bibinfo
  {volume} {10}},\ \bibinfo {pages} {604} (\bibinfo {year} {2020})}\BibitemShut
  {NoStop}%
\bibitem [{\citenamefont {Park}\ and\ \citenamefont
  {Lavrentovich}(2012)}]{park:lyotropic}%
  \BibitemOpen
  \bibfield  {author} {\bibinfo {author} {\bibfnamefont {H.-S.}\ \bibnamefont
  {Park}}\ and\ \bibinfo {author} {\bibfnamefont {O.~D.}\ \bibnamefont
  {Lavrentovich}},\ }\bibfield  {title} {\bibinfo {title} {Lyotropic chromonic
  liquid crystals: {E}merging applications},\ }in\ \href@noop {} {\emph
  {\bibinfo {booktitle} {Liquid Crystals Beyond Displays: {C}hemistry, Physics,
  and Applications}}},\ \bibinfo {editor} {edited by\ \bibinfo {editor}
  {\bibfnamefont {Q.}~\bibnamefont {Li}}}\ (\bibinfo  {publisher} {John Wiley
  \& Sons},\ \bibinfo {address} {Hoboken, New Jersey},\ \bibinfo {year}
  {2012})\ pp.\ \bibinfo {pages} {449--48}\BibitemShut {NoStop}%
\bibitem [{\citenamefont {Shiyanovskii}\ \emph {et~al.}(2005)\citenamefont
  {Shiyanovskii}, \citenamefont {Schneider}, \citenamefont {Smalyukh},
  \citenamefont {Ishikawa}, \citenamefont {Niehaus}, \citenamefont {Doane},
  \citenamefont {Woolverton},\ and\ \citenamefont
  {Lavrentovich}}]{shiyanovskii:real-time}%
  \BibitemOpen
  \bibfield  {author} {\bibinfo {author} {\bibfnamefont {S.~V.}\ \bibnamefont
  {Shiyanovskii}}, \bibinfo {author} {\bibfnamefont {T.}~\bibnamefont
  {Schneider}}, \bibinfo {author} {\bibfnamefont {I.~I.}\ \bibnamefont
  {Smalyukh}}, \bibinfo {author} {\bibfnamefont {T.}~\bibnamefont {Ishikawa}},
  \bibinfo {author} {\bibfnamefont {G.~D.}\ \bibnamefont {Niehaus}}, \bibinfo
  {author} {\bibfnamefont {K.~J.}\ \bibnamefont {Doane}}, \bibinfo {author}
  {\bibfnamefont {C.~J.}\ \bibnamefont {Woolverton}},\ and\ \bibinfo {author}
  {\bibfnamefont {O.~D.}\ \bibnamefont {Lavrentovich}},\ }\bibfield  {title}
  {\bibinfo {title} {Real-time microbe detection based on director distortions
  around growing immune complexes in lyotropic chromonic liquid crystals},\
  }\href {https://doi.org/10.1103/PhysRevE.71.020702} {\bibfield  {journal}
  {\bibinfo  {journal} {Phys. Rev. E}\ }\textbf {\bibinfo {volume} {71}},\
  \bibinfo {pages} {020702} (\bibinfo {year} {2005})}\BibitemShut {NoStop}%
\bibitem [{\citenamefont {Woolverton}\ \emph {et~al.}(2005)\citenamefont
  {Woolverton}, \citenamefont {Gustely}, \citenamefont {Li},\ and\
  \citenamefont {Lavrentovich}}]{woolverton:liquid}%
  \BibitemOpen
  \bibfield  {author} {\bibinfo {author} {\bibfnamefont {C.~J.}\ \bibnamefont
  {Woolverton}}, \bibinfo {author} {\bibfnamefont {E.}~\bibnamefont {Gustely}},
  \bibinfo {author} {\bibfnamefont {L.}~\bibnamefont {Li}},\ and\ \bibinfo
  {author} {\bibfnamefont {O.~D.}\ \bibnamefont {Lavrentovich}},\ }\bibfield
  {title} {\bibinfo {title} {Liquid crystal effects on bacterial viability},\
  }\href {https://doi.org/10.1080/02678290500074822} {\bibfield  {journal}
  {\bibinfo  {journal} {Liq. Cryst.}\ }\textbf {\bibinfo {volume} {32}},\
  \bibinfo {pages} {417} (\bibinfo {year} {2005})}\BibitemShut {NoStop}%
\bibitem [{\citenamefont {Shaban}\ \emph {et~al.}(2021)\citenamefont {Shaban},
  \citenamefont {Mon-Juan},\ and\ \citenamefont {Lee}}]{shaban:label-free}%
  \BibitemOpen
  \bibfield  {author} {\bibinfo {author} {\bibfnamefont {H.}~\bibnamefont
  {Shaban}}, \bibinfo {author} {\bibfnamefont {L.}~\bibnamefont {Mon-Juan}},\
  and\ \bibinfo {author} {\bibfnamefont {W.}~\bibnamefont {Lee}},\ }\bibfield
  {title} {\bibinfo {title} {Label-free detection and spectrometrically
  quantitative analysis of the cancer biomarker {CA125} based on lyotropic
  chromonic liquid crystal},\ }\href
  {https://doi.org/https://doi.org/10.3390/bios11080271} {\bibfield  {journal}
  {\bibinfo  {journal} {Biosensors}\ }\textbf {\bibinfo {volume} {11}},\
  \bibinfo {pages} {271} (\bibinfo {year} {2021})}\BibitemShut {NoStop}%
\bibitem [{\citenamefont {Yi}\ and\ \citenamefont
  {Clark}(2013)}]{yi:orientation}%
  \BibitemOpen
  \bibfield  {author} {\bibinfo {author} {\bibfnamefont {Y.}~\bibnamefont
  {Yi}}\ and\ \bibinfo {author} {\bibfnamefont {N.~A.}\ \bibnamefont {Clark}},\
  }\bibfield  {title} {\bibinfo {title} {Orientation of chromonic liquid
  crystals by topographic linear channels: multi-stable alignment and tactoid
  structure},\ }\href {https://doi.org/10.1080/02678292.2013.839831} {\bibfield
   {journal} {\bibinfo  {journal} {Liq. Cryst.}\ }\textbf {\bibinfo {volume}
  {40}},\ \bibinfo {pages} {1736} (\bibinfo {year} {2013})}\BibitemShut
  {NoStop}%
\bibitem [{\citenamefont {Mcguire}\ \emph {et~al.}(2014)\citenamefont
  {Mcguire}, \citenamefont {Yi},\ and\ \citenamefont
  {Clark}}]{mcguire:orthogonal}%
  \BibitemOpen
  \bibfield  {author} {\bibinfo {author} {\bibfnamefont {A.}~\bibnamefont
  {Mcguire}}, \bibinfo {author} {\bibfnamefont {Y.}~\bibnamefont {Yi}},\ and\
  \bibinfo {author} {\bibfnamefont {N.~A.}\ \bibnamefont {Clark}},\ }\bibfield
  {title} {\bibinfo {title} {Orthogonal orientation of chromonic liquid
  crystals by rubbed polyamide films},\ }\href
  {https://doi.org/https://doi.org/10.1002/cphc.201301040} {\bibfield
  {journal} {\bibinfo  {journal} {Chem. Phys. Chem.}\ }\textbf {\bibinfo
  {volume} {15}},\ \bibinfo {pages} {1376} (\bibinfo {year}
  {2014})}\BibitemShut {NoStop}%
\bibitem [{\citenamefont {Paparini}\ and\ \citenamefont
  {Virga}(2021{\natexlab{a}})}]{paparini:shape}%
  \BibitemOpen
  \bibfield  {author} {\bibinfo {author} {\bibfnamefont {S.}~\bibnamefont
  {Paparini}}\ and\ \bibinfo {author} {\bibfnamefont {E.~G.}\ \bibnamefont
  {Virga}},\ }\bibfield  {title} {\bibinfo {title} {Shape bistability in 2{D}
  chromonic droplets},\ }\href {https://doi.org/10.1088/1361-648x/ac2645}
  {\bibfield  {journal} {\bibinfo  {journal} {J. Phys.: Condens. Matter}\
  }\textbf {\bibinfo {volume} {33}},\ \bibinfo {pages} {495101} (\bibinfo
  {year} {2021}{\natexlab{a}})}\BibitemShut {NoStop}%
\bibitem [{\citenamefont {Nayani}\ \emph {et~al.}(2015)\citenamefont {Nayani},
  \citenamefont {Chang}, \citenamefont {Fu}, \citenamefont {Ellis},
  \citenamefont {Fernandez-Nieves}, \citenamefont {Park},\ and\ \citenamefont
  {Srinivasarao}}]{nayani:spontaneous}%
  \BibitemOpen
  \bibfield  {author} {\bibinfo {author} {\bibfnamefont {K.}~\bibnamefont
  {Nayani}}, \bibinfo {author} {\bibfnamefont {R.}~\bibnamefont {Chang}},
  \bibinfo {author} {\bibfnamefont {J.}~\bibnamefont {Fu}}, \bibinfo {author}
  {\bibfnamefont {P.~W.}\ \bibnamefont {Ellis}}, \bibinfo {author}
  {\bibfnamefont {A.}~\bibnamefont {Fernandez-Nieves}}, \bibinfo {author}
  {\bibfnamefont {J.~O.}\ \bibnamefont {Park}},\ and\ \bibinfo {author}
  {\bibfnamefont {M.}~\bibnamefont {Srinivasarao}},\ }\bibfield  {title}
  {\bibinfo {title} {Spontaneous emergence of chirality in achiral lyotropic
  chromonic liquid crystals confined to cylinders},\ }\href
  {https://doi.org/https://doi.org/10.1038/ncomms9067} {\bibfield  {journal}
  {\bibinfo  {journal} {Nat. Commun.}\ }\textbf {\bibinfo {volume} {6}},\
  \bibinfo {pages} {8067} (\bibinfo {year} {2015})}\BibitemShut {NoStop}%
\bibitem [{\citenamefont {Davidson}\ \emph
  {et~al.}(2015{\natexlab{a}})\citenamefont {Davidson}, \citenamefont {Kang},
  \citenamefont {Jeong}, \citenamefont {Still}, \citenamefont {Collings},
  \citenamefont {Lubensky},\ and\ \citenamefont {Yodh}}]{davidson:chiral}%
  \BibitemOpen
  \bibfield  {author} {\bibinfo {author} {\bibfnamefont {Z.~S.}\ \bibnamefont
  {Davidson}}, \bibinfo {author} {\bibfnamefont {L.}~\bibnamefont {Kang}},
  \bibinfo {author} {\bibfnamefont {J.}~\bibnamefont {Jeong}}, \bibinfo
  {author} {\bibfnamefont {T.}~\bibnamefont {Still}}, \bibinfo {author}
  {\bibfnamefont {P.~J.}\ \bibnamefont {Collings}}, \bibinfo {author}
  {\bibfnamefont {T.~C.}\ \bibnamefont {Lubensky}},\ and\ \bibinfo {author}
  {\bibfnamefont {A.~G.}\ \bibnamefont {Yodh}},\ }\bibfield  {title} {\bibinfo
  {title} {Chiral structures and defects of lyotropic chromonic liquid crystals
  induced by saddle-splay elasticity},\ }\href
  {https://doi.org/10.1103/PhysRevE.91.050501} {\bibfield  {journal} {\bibinfo
  {journal} {Phys. Rev. E}\ }\textbf {\bibinfo {volume} {91}},\ \bibinfo
  {pages} {050501({R})} (\bibinfo {year} {2015}{\natexlab{a}})},\ \bibinfo
  {note} {see also {E}rratum \cite{davidson:erratum} and {S}upplementary
  {I}nformation
  \url{https://journals.aps.org/pre/supplemental/10.1103/PhysRevE.91.050501/Supplementary_Info_Planar_Davidson_et_al.pdf}.}\BibitemShut
  {Stop}%
\bibitem [{\citenamefont {Fu}\ \emph {et~al.}(2017)\citenamefont {Fu},
  \citenamefont {Nayani}, \citenamefont {Park},\ and\ \citenamefont
  {Srinivasarao}}]{fu:spontaneous}%
  \BibitemOpen
  \bibfield  {author} {\bibinfo {author} {\bibfnamefont {J.}~\bibnamefont
  {Fu}}, \bibinfo {author} {\bibfnamefont {K.}~\bibnamefont {Nayani}}, \bibinfo
  {author} {\bibfnamefont {J.}~\bibnamefont {Park}},\ and\ \bibinfo {author}
  {\bibfnamefont {M.}~\bibnamefont {Srinivasarao}},\ }\bibfield  {title}
  {\bibinfo {title} {Spontaneous emergence of twist and formation of monodomain
  in lyotropic chromonic liquid crystals confined to capillaries},\ }\href
  {https://doi.org/10.1038/am.2017.84} {\bibfield  {journal} {\bibinfo
  {journal} {NPG Asia Mater.}\ }\textbf {\bibinfo {volume} {9}},\ \bibinfo
  {pages} {e393} (\bibinfo {year} {2017})}\BibitemShut {NoStop}%
\bibitem [{\citenamefont {Machon}\ and\ \citenamefont
  {Alexander}(2016)}]{machon:umbilic}%
  \BibitemOpen
  \bibfield  {author} {\bibinfo {author} {\bibfnamefont {T.}~\bibnamefont
  {Machon}}\ and\ \bibinfo {author} {\bibfnamefont {G.~P.}\ \bibnamefont
  {Alexander}},\ }\bibfield  {title} {\bibinfo {title} {Umbilic lines in
  orientational order},\ }\href {https://doi.org/10.1103/PhysRevX.6.011033}
  {\bibfield  {journal} {\bibinfo  {journal} {Phys. Rev. X}\ }\textbf {\bibinfo
  {volume} {6}},\ \bibinfo {pages} {011033} (\bibinfo {year}
  {2016})}\BibitemShut {NoStop}%
\bibitem [{\citenamefont {Selinger}(2018)}]{selinger:interpretation}%
  \BibitemOpen
  \bibfield  {author} {\bibinfo {author} {\bibfnamefont {J.~V.}\ \bibnamefont
  {Selinger}},\ }\bibfield  {title} {\bibinfo {title} {Interpretation of
  saddle-splay and the {O}seen-{F}rank free energy in liquid crystals},\ }\href
  {https://doi.org/10.1080/21680396.2019.1581103} {\bibfield  {journal}
  {\bibinfo  {journal} {Liq. Cryst. Rev.}\ }\textbf {\bibinfo {volume} {6}},\
  \bibinfo {pages} {129} (\bibinfo {year} {2018})}\BibitemShut {NoStop}%
\bibitem [{\citenamefont {Pedrini}\ and\ \citenamefont
  {Virga}(2020)}]{pedrini:liquid}%
  \BibitemOpen
  \bibfield  {author} {\bibinfo {author} {\bibfnamefont {A.}~\bibnamefont
  {Pedrini}}\ and\ \bibinfo {author} {\bibfnamefont {E.~G.}\ \bibnamefont
  {Virga}},\ }\bibfield  {title} {\bibinfo {title} {Liquid crystal distortions
  revealed by an octupolar tensor},\ }\href
  {https://doi.org/10.1103/PhysRevE.101.012703} {\bibfield  {journal} {\bibinfo
   {journal} {Phys. Rev. E}\ }\textbf {\bibinfo {volume} {101}},\ \bibinfo
  {pages} {012703} (\bibinfo {year} {2020})}\BibitemShut {NoStop}%
\bibitem [{\citenamefont {Selinger}(2022)}]{selinger:director}%
  \BibitemOpen
  \bibfield  {author} {\bibinfo {author} {\bibfnamefont {J.~V.}\ \bibnamefont
  {Selinger}},\ }\bibfield  {title} {\bibinfo {title} {Director deformations,
  geometric frustration, and modulated phases in liquid crystals},\ }\href
  {https://doi.org/10.1146/annurev-conmatphys-031620-105712} {\bibfield
  {journal} {\bibinfo  {journal} {Ann. Rev. Condens. Matter Phys.}\ }\textbf
  {\bibinfo {volume} {13}},\ \bibinfo {pages} {49} (\bibinfo {year}
  {2022})}\BibitemShut {NoStop}%
\bibitem [{\citenamefont {Virga}(2019)}]{virga:uniform}%
  \BibitemOpen
  \bibfield  {author} {\bibinfo {author} {\bibfnamefont {E.~G.}\ \bibnamefont
  {Virga}},\ }\bibfield  {title} {\bibinfo {title} {Uniform distortions and
  generalized elasticity of liquid crystals},\ }\href
  {https://doi.org/10.1103/PhysRevE.100.052701} {\bibfield  {journal} {\bibinfo
   {journal} {Phys. Rev. E}\ }\textbf {\bibinfo {volume} {100}},\ \bibinfo
  {pages} {052701} (\bibinfo {year} {2019})}\BibitemShut {NoStop}%
\bibitem [{\citenamefont {Ericksen}(1966)}]{ericksen:inequalities}%
  \BibitemOpen
  \bibfield  {author} {\bibinfo {author} {\bibfnamefont {J.~L.}\ \bibnamefont
  {Ericksen}},\ }\bibfield  {title} {\bibinfo {title} {Inequalities in liquid
  crystal theory},\ }\href {https://doi.org/10.1063/1.1761821} {\bibfield
  {journal} {\bibinfo  {journal} {Phys. Fluids}\ }\textbf {\bibinfo {volume}
  {9}},\ \bibinfo {pages} {1205} (\bibinfo {year} {1966})}\BibitemShut
  {NoStop}%
\bibitem [{\citenamefont {Paparini}\ and\ \citenamefont
  {Virga}(2022{\natexlab{a}})}]{paparini:paradoxes}%
  \BibitemOpen
  \bibfield  {author} {\bibinfo {author} {\bibfnamefont {S.}~\bibnamefont
  {Paparini}}\ and\ \bibinfo {author} {\bibfnamefont {E.~G.}\ \bibnamefont
  {Virga}},\ }\bibfield  {title} {\bibinfo {title} {Paradoxes for chromonic
  liquid crystal droplets},\ }\href
  {https://doi.org/10.1103/PhysRevE.106.044703} {\bibfield  {journal} {\bibinfo
   {journal} {Phys. Rev. E}\ }\textbf {\bibinfo {volume} {106}},\ \bibinfo
  {pages} {044703} (\bibinfo {year} {2022}{\natexlab{a}})}\BibitemShut
  {NoStop}%
\bibitem [{\citenamefont {Paparini}\ and\ \citenamefont
  {Virga}(2022{\natexlab{b}})}]{paparini:stability}%
  \BibitemOpen
  \bibfield  {author} {\bibinfo {author} {\bibfnamefont {S.}~\bibnamefont
  {Paparini}}\ and\ \bibinfo {author} {\bibfnamefont {E.~G.}\ \bibnamefont
  {Virga}},\ }\bibfield  {title} {\bibinfo {title} {Stability against the odds:
  the case of chromonic liquid crystals},\ }\href
  {https://doi.org/https://doi.org/10.1007/s00332-022-09833-6} {\bibfield
  {journal} {\bibinfo  {journal} {J. Nonlinear Sci.}\ }\textbf {\bibinfo
  {volume} {32}},\ \bibinfo {pages} {74} (\bibinfo {year}
  {2022}{\natexlab{b}})}\BibitemShut {NoStop}%
\bibitem [{\citenamefont {Long}\ and\ \citenamefont
  {Selinger}(2023)}]{long:violation}%
  \BibitemOpen
  \bibfield  {author} {\bibinfo {author} {\bibfnamefont {C.}~\bibnamefont
  {Long}}\ and\ \bibinfo {author} {\bibfnamefont {J.~V.}\ \bibnamefont
  {Selinger}},\ }\bibfield  {title} {\bibinfo {title} {Violation of {E}ricksen
  inequalities in lyotropic chromonic liquid crystals},\ }\href
  {https://doi.org/https://doi.org/10.1007/s10659-022-09899-z} {\bibfield
  {journal} {\bibinfo  {journal} {J. Elast.}\ }\textbf {\bibinfo {volume}
  {153}},\ \bibinfo {pages} {599} (\bibinfo {year} {2023})}\BibitemShut
  {NoStop}%
\bibitem [{\citenamefont {Paparini}\ and\ \citenamefont
  {Virga}(2023{\natexlab{a}})}]{paparini:elastic}%
  \BibitemOpen
  \bibfield  {author} {\bibinfo {author} {\bibfnamefont {S.}~\bibnamefont
  {Paparini}}\ and\ \bibinfo {author} {\bibfnamefont {E.~G.}\ \bibnamefont
  {Virga}},\ }\bibfield  {title} {\bibinfo {title} {An elastic quartic twist
  theory for chromonic liquid crystals},\ }\href
  {https://doi.org/https://doi.org/10.1007/s10659-022-09983-4} {\bibfield
  {journal} {\bibinfo  {journal} {J. Elast.}\ } (\bibinfo {year}
  {2023}{\natexlab{a}})}\BibitemShut {NoStop}%
\bibitem [{\citenamefont {Paparini}\ and\ \citenamefont
  {Virga}(2023{\natexlab{b}})}]{paparini:spiralling}%
  \BibitemOpen
  \bibfield  {author} {\bibinfo {author} {\bibfnamefont {S.}~\bibnamefont
  {Paparini}}\ and\ \bibinfo {author} {\bibfnamefont {E.~G.}\ \bibnamefont
  {Virga}},\ }\bibfield  {title} {\bibinfo {title} {Spiralling defect cores in
  chromonic hedgehogs},\ }\href {https://doi.org/10.1080/02678292.2023.2190626}
  {\bibfield  {journal} {\bibinfo  {journal} {Liq. Cryst.}\ } (\bibinfo {year}
  {2023}{\natexlab{b}})}\BibitemShut {NoStop}%
\bibitem [{\citenamefont {Pedrini}\ and\ \citenamefont
  {Virga}(2023)}]{pedrini:relieving}%
  \BibitemOpen
  \bibfield  {author} {\bibinfo {author} {\bibfnamefont {A.}~\bibnamefont
  {Pedrini}}\ and\ \bibinfo {author} {\bibfnamefont {E.~G.}\ \bibnamefont
  {Virga}},\ }\bibfield  {title} {\bibinfo {title} {Relieving nematic geometric
  frustration in the plane},\ }\href {https://doi.org/10.1088/1751-8121/acd890}
  {\bibfield  {journal} {\bibinfo  {journal} {J. Phys. A: Math. Theor.}\
  }\textbf {\bibinfo {volume} {56}},\ \bibinfo {pages} {265202} (\bibinfo
  {year} {2023})}\BibitemShut {NoStop}%
\bibitem [{\citenamefont {Rapini}\ and\ \citenamefont
  {Papoular}(1969)}]{rapini:distortion}%
  \BibitemOpen
  \bibfield  {author} {\bibinfo {author} {\bibfnamefont {A.}~\bibnamefont
  {Rapini}}\ and\ \bibinfo {author} {\bibfnamefont {M.}~\bibnamefont
  {Papoular}},\ }\bibfield  {title} {\bibinfo {title} {Distorsion d'une lamelle
  m{\'e}matique sous champ magn{\'e}tique conditions d'ancrage aux parois},\
  }\href@noop {} {\bibfield  {journal} {\bibinfo  {journal} {J. Phys. Colloq.}\
  }\textbf {\bibinfo {volume} {30}},\ \bibinfo {pages} {C4.54} (\bibinfo {year}
  {1969})},\ \bibinfo {note} {available from
  \url{https://hal.archives-ouvertes.fr/jpa-00213715/document}}\BibitemShut
  {NoStop}%
\bibitem [{\citenamefont {Yoon}\ \emph {et~al.}(2010)\citenamefont {Yoon},
  \citenamefont {Deb}, \citenamefont {Chen}, \citenamefont {K\"orblova},
  \citenamefont {Shao}, \citenamefont {Ishikawa}, \citenamefont {Rao},
  \citenamefont {Walba}, \citenamefont {Smalyukh},\ and\ \citenamefont
  {Clark}}]{yoon:organization}%
  \BibitemOpen
  \bibfield  {author} {\bibinfo {author} {\bibfnamefont {D.~K.}\ \bibnamefont
  {Yoon}}, \bibinfo {author} {\bibfnamefont {R.}~\bibnamefont {Deb}}, \bibinfo
  {author} {\bibfnamefont {D.}~\bibnamefont {Chen}}, \bibinfo {author}
  {\bibfnamefont {E.}~\bibnamefont {K\"orblova}}, \bibinfo {author}
  {\bibfnamefont {R.}~\bibnamefont {Shao}}, \bibinfo {author} {\bibfnamefont
  {K.}~\bibnamefont {Ishikawa}}, \bibinfo {author} {\bibfnamefont {N.~V.~S.}\
  \bibnamefont {Rao}}, \bibinfo {author} {\bibfnamefont {D.~M.}\ \bibnamefont
  {Walba}}, \bibinfo {author} {\bibfnamefont {I.~I.}\ \bibnamefont
  {Smalyukh}},\ and\ \bibinfo {author} {\bibfnamefont {N.~A.}\ \bibnamefont
  {Clark}},\ }\bibfield  {title} {\bibinfo {title} {Organization of the
  polarization splay modulated smectic liquid crystal phase by topographic
  confinement},\ }\href {https://doi.org/10.1073/pnas.1014593107} {\bibfield
  {journal} {\bibinfo  {journal} {Proc. Natl. Acad. Sci. USA}\ }\textbf
  {\bibinfo {volume} {107}},\ \bibinfo {pages} {21311} (\bibinfo {year}
  {2010})}\BibitemShut {NoStop}%
\bibitem [{\citenamefont {Kim}\ \emph {et~al.}(2000)\citenamefont {Kim},
  \citenamefont {Kim}, \citenamefont {Fukuda},\ and\ \citenamefont
  {Matsuda}}]{kim:alignment}%
  \BibitemOpen
  \bibfield  {author} {\bibinfo {author} {\bibfnamefont {M.-H.}\ \bibnamefont
  {Kim}}, \bibinfo {author} {\bibfnamefont {J.-D.}\ \bibnamefont {Kim}},
  \bibinfo {author} {\bibfnamefont {T.}~\bibnamefont {Fukuda}},\ and\ \bibinfo
  {author} {\bibfnamefont {H.}~\bibnamefont {Matsuda}},\ }\bibfield  {title}
  {\bibinfo {title} {Alignment control of liquid crystals on surface relief
  gratings},\ }\href {https://doi.org/10.1080/026782900750037194} {\bibfield
  {journal} {\bibinfo  {journal} {Liq. Cryst.}\ }\textbf {\bibinfo {volume}
  {27}},\ \bibinfo {pages} {1633} (\bibinfo {year} {2000})}\BibitemShut
  {NoStop}%
\bibitem [{\citenamefont {Behdani}\ \emph {et~al.}(2003)\citenamefont
  {Behdani}, \citenamefont {Keshmiri}, \citenamefont {Soria}, \citenamefont
  {Bader}, \citenamefont {Ihlemann}, \citenamefont {Marowsky},\ and\
  \citenamefont {Rasing}}]{behdani:alignment}%
  \BibitemOpen
  \bibfield  {author} {\bibinfo {author} {\bibfnamefont {M.}~\bibnamefont
  {Behdani}}, \bibinfo {author} {\bibfnamefont {S.~H.}\ \bibnamefont
  {Keshmiri}}, \bibinfo {author} {\bibfnamefont {S.}~\bibnamefont {Soria}},
  \bibinfo {author} {\bibfnamefont {M.~A.}\ \bibnamefont {Bader}}, \bibinfo
  {author} {\bibfnamefont {J.}~\bibnamefont {Ihlemann}}, \bibinfo {author}
  {\bibfnamefont {G.}~\bibnamefont {Marowsky}},\ and\ \bibinfo {author}
  {\bibfnamefont {T.}~\bibnamefont {Rasing}},\ }\bibfield  {title} {\bibinfo
  {title} {Alignment of liquid crystals with periodic submicron structures
  ablated in polymeric and indium tin oxide surfaces},\ }\href
  {https://doi.org/10.1063/1.1565703} {\bibfield  {journal} {\bibinfo
  {journal} {Appl. Phys. Lett.}\ }\textbf {\bibinfo {volume} {82}},\ \bibinfo
  {pages} {2553} (\bibinfo {year} {2003})}\BibitemShut {NoStop}%
\bibitem [{\citenamefont {Geng}\ and\ \citenamefont
  {Lin}(2022)}]{geng:two-dimensional}%
  \BibitemOpen
  \bibfield  {author} {\bibinfo {author} {\bibfnamefont {Z.}~\bibnamefont
  {Geng}}\ and\ \bibinfo {author} {\bibfnamefont {F.}~\bibnamefont {Lin}},\
  }\bibfield  {title} {\bibinfo {title} {The two-dimensional liquid crystal
  droplet problem with a tangential boundary condition},\ }\href
  {https://doi.org/https://doi.org/10.1007/s00205-021-01733-5} {\bibfield
  {journal} {\bibinfo  {journal} {Arch. Rational Mech. Anal.}\ }\textbf
  {\bibinfo {volume} {243}},\ \bibinfo {pages} {1181} (\bibinfo {year}
  {2022})}\BibitemShut {NoStop}%
\bibitem [{\citenamefont {Kaznacheev}\ \emph {et~al.}(2002)\citenamefont
  {Kaznacheev}, \citenamefont {Bogdanov},\ and\ \citenamefont
  {A.Taraskin}}]{kaznacheev:nature}%
  \BibitemOpen
  \bibfield  {author} {\bibinfo {author} {\bibfnamefont {A.~V.}\ \bibnamefont
  {Kaznacheev}}, \bibinfo {author} {\bibfnamefont {M.~M.}\ \bibnamefont
  {Bogdanov}},\ and\ \bibinfo {author} {\bibfnamefont {S.}~\bibnamefont
  {A.Taraskin}},\ }\bibfield  {title} {\bibinfo {title} {The nature of prolate
  shape of tactoids in lyotropic inorganic liquid crystals},\ }\href
  {https://doi.org/https://doi.org/10.1134/1.1499901} {\bibfield  {journal}
  {\bibinfo  {journal} {J. Exp. Theor. Phys.}\ }\textbf {\bibinfo {volume}
  {95}},\ \bibinfo {pages} {57} (\bibinfo {year} {2002})}\BibitemShut {NoStop}%
\bibitem [{\citenamefont {Kaznacheev}\ \emph {et~al.}(2003)\citenamefont
  {Kaznacheev}, \citenamefont {Bogdanov},\ and\ \citenamefont
  {Sonin}}]{kaznacheev:influence}%
  \BibitemOpen
  \bibfield  {author} {\bibinfo {author} {\bibfnamefont {A.~V.}\ \bibnamefont
  {Kaznacheev}}, \bibinfo {author} {\bibfnamefont {M.~M.}\ \bibnamefont
  {Bogdanov}},\ and\ \bibinfo {author} {\bibfnamefont {A.~S.}\ \bibnamefont
  {Sonin}},\ }\bibfield  {title} {\bibinfo {title} {The influence of anchoring
  energy on the prolate shape of tactoids in lyotropic inorganic liquid
  crystals},\ }\href {https://doi.org/https://doi.org/10.1134/1.1641899}
  {\bibfield  {journal} {\bibinfo  {journal} {J. Exp. Theor. Phys.}\ }\textbf
  {\bibinfo {volume} {97}},\ \bibinfo {pages} {1159} (\bibinfo {year}
  {2003})}\BibitemShut {NoStop}%
\bibitem [{\citenamefont {Prinsen}\ and\ \citenamefont {van~der
  Schoot}(2003)}]{prinsen:shape}%
  \BibitemOpen
  \bibfield  {author} {\bibinfo {author} {\bibfnamefont {P.}~\bibnamefont
  {Prinsen}}\ and\ \bibinfo {author} {\bibfnamefont {P.}~\bibnamefont {van~der
  Schoot}},\ }\bibfield  {title} {\bibinfo {title} {Shape and director-field
  transformation of tactoids},\ }\href
  {https://doi.org/10.1103/PhysRevE.68.021701} {\bibfield  {journal} {\bibinfo
  {journal} {Phys. Rev. E}\ }\textbf {\bibinfo {volume} {68}},\ \bibinfo
  {pages} {021701} (\bibinfo {year} {2003})}\BibitemShut {NoStop}%
\bibitem [{\citenamefont {Prinsen}\ and\ \citenamefont {van~der
  Schoot}(2004)}]{prinsen:parity}%
  \BibitemOpen
  \bibfield  {author} {\bibinfo {author} {\bibfnamefont {P.}~\bibnamefont
  {Prinsen}}\ and\ \bibinfo {author} {\bibfnamefont {P.}~\bibnamefont {van~der
  Schoot}},\ }\bibfield  {title} {\bibinfo {title} {Parity breaking in nematic
  tactoids},\ }\href {https://doi.org/10.1088/0953-8984/16/49/003} {\bibfield
  {journal} {\bibinfo  {journal} {J. Phys.: Condens. Matter}\ }\textbf
  {\bibinfo {volume} {16}},\ \bibinfo {pages} {8835} (\bibinfo {year}
  {2004})}\BibitemShut {NoStop}%
\bibitem [{\citenamefont {Prinsen}\ and\ \citenamefont {{van der
  Schoot}}(2004)}]{prinsen:continuous}%
  \BibitemOpen
  \bibfield  {author} {\bibinfo {author} {\bibfnamefont {P.}~\bibnamefont
  {Prinsen}}\ and\ \bibinfo {author} {\bibfnamefont {P.}~\bibnamefont {{van der
  Schoot}}},\ }\bibfield  {title} {\bibinfo {title} {Continuous director-field
  transformation of nematic tactoids},\ }\href
  {https://doi.org/https://doi.org/10.1140/epje/e2004-00038-y} {\bibfield
  {journal} {\bibinfo  {journal} {Euro. Phys. J. E}\ }\textbf {\bibinfo
  {volume} {13}},\ \bibinfo {pages} {35} (\bibinfo {year} {2004})}\BibitemShut
  {NoStop}%
\bibitem [{\citenamefont {Puech}\ \emph {et~al.}(2010)\citenamefont {Puech},
  \citenamefont {Grelet}, \citenamefont {Poulin}, \citenamefont {Blanc},\ and\
  \citenamefont {van~der Schoot}}]{puech:nematic}%
  \BibitemOpen
  \bibfield  {author} {\bibinfo {author} {\bibfnamefont {N.}~\bibnamefont
  {Puech}}, \bibinfo {author} {\bibfnamefont {E.}~\bibnamefont {Grelet}},
  \bibinfo {author} {\bibfnamefont {P.}~\bibnamefont {Poulin}}, \bibinfo
  {author} {\bibfnamefont {C.}~\bibnamefont {Blanc}},\ and\ \bibinfo {author}
  {\bibfnamefont {P.}~\bibnamefont {van~der Schoot}},\ }\bibfield  {title}
  {\bibinfo {title} {Nematic droplets in aqueous dispersions of carbon
  nanotubes},\ }\href {https://doi.org/10.1103/PhysRevE.82.020702} {\bibfield
  {journal} {\bibinfo  {journal} {Phys. Rev. E}\ }\textbf {\bibinfo {volume}
  {82}},\ \bibinfo {pages} {020702} (\bibinfo {year} {2010})}\BibitemShut
  {NoStop}%
\bibitem [{\citenamefont {Verhoeff}\ \emph {et~al.}(2011)\citenamefont
  {Verhoeff}, \citenamefont {Bakelaar}, \citenamefont {Otten}, \citenamefont
  {{van der Schoot}},\ and\ \citenamefont {Lekkerkerker}}]{verhoeff:tactoids}%
  \BibitemOpen
  \bibfield  {author} {\bibinfo {author} {\bibfnamefont {A.~A.}\ \bibnamefont
  {Verhoeff}}, \bibinfo {author} {\bibfnamefont {I.~A.}\ \bibnamefont
  {Bakelaar}}, \bibinfo {author} {\bibfnamefont {R.~H.~J.}\ \bibnamefont
  {Otten}}, \bibinfo {author} {\bibfnamefont {P.}~\bibnamefont {{van der
  Schoot}}},\ and\ \bibinfo {author} {\bibfnamefont {H.~N.~W.}\ \bibnamefont
  {Lekkerkerker}},\ }\bibfield  {title} {\bibinfo {title} {Tactoids of
  plate-like particles: Size, shape, and director field},\ }\href
  {https://doi.org/https://doi.org/10.1021/la104128m} {\bibfield  {journal}
  {\bibinfo  {journal} {Langmuir}\ }\textbf {\bibinfo {volume} {27}},\ \bibinfo
  {pages} {116} (\bibinfo {year} {2011})}\BibitemShut {NoStop}%
\bibitem [{\citenamefont {Kleman}\ and\ \citenamefont
  {Lavrentovich}(2003)}]{kleman:soft}%
  \BibitemOpen
  \bibfield  {author} {\bibinfo {author} {\bibfnamefont {M.}~\bibnamefont
  {Kleman}}\ and\ \bibinfo {author} {\bibfnamefont {O.~D.}\ \bibnamefont
  {Lavrentovich}},\ }\href@noop {} {\emph {\bibinfo {title} {Soft Matter
  Physics: {A}n Introduction}}},\ Partially {O}rdered {S}ystems\ (\bibinfo
  {publisher} {Springer-Verlag},\ \bibinfo {address} {New York},\ \bibinfo
  {year} {2003})\BibitemShut {NoStop}%
\bibitem [{\citenamefont {{de~G}ennes}\ and\ \citenamefont
  {Prost}(1993)}]{degennes:physics}%
  \BibitemOpen
  \bibfield  {author} {\bibinfo {author} {\bibfnamefont {P.~G.}\ \bibnamefont
  {{de~G}ennes}}\ and\ \bibinfo {author} {\bibfnamefont {J.}~\bibnamefont
  {Prost}},\ }\href@noop {} {\emph {\bibinfo {title} {The Physics of Liquid
  Crystals}}},\ \bibinfo {edition} {2nd}\ ed.,\ \bibinfo {series} {The
  {I}nternational {S}eries of {M}onographs on {P}hysics}, Vol.~\bibinfo
  {volume} {83}\ (\bibinfo  {publisher} {Clarendon Press},\ \bibinfo {address}
  {Oxford},\ \bibinfo {year} {1993})\BibitemShut {NoStop}%
\bibitem [{\citenamefont {Zhang}\ and\ \citenamefont
  {Kitzerow}(2016)}]{zhang:influence}%
  \BibitemOpen
  \bibfield  {author} {\bibinfo {author} {\bibfnamefont {B.}~\bibnamefont
  {Zhang}}\ and\ \bibinfo {author} {\bibfnamefont {H.-S.}\ \bibnamefont
  {Kitzerow}},\ }\bibfield  {title} {\bibinfo {title} {Influence of proton and
  salt concentration on the chromonic liquid crystal phase diagram of disodium
  cromoglycate solutions: {P}rospects and limitations of a host for {DNA}
  nanostructures},\ }\href {https://doi.org/10.1021/acs.jpcb.6b01644}
  {\bibfield  {journal} {\bibinfo  {journal} {J. Phys. Chem. B}\ }\textbf
  {\bibinfo {volume} {120}},\ \bibinfo {pages} {3250} (\bibinfo {year}
  {2016})}\BibitemShut {NoStop}%
\bibitem [{\citenamefont {Zhou}\ \emph {et~al.}(2014)\citenamefont {Zhou},
  \citenamefont {Neupane}, \citenamefont {Nastishin}, \citenamefont {Baldwin},
  \citenamefont {Shiyanovskii}, \citenamefont {Lavrentovich},\ and\
  \citenamefont {Sprunt}}]{zhou:elasticity_2014}%
  \BibitemOpen
  \bibfield  {author} {\bibinfo {author} {\bibfnamefont {S.}~\bibnamefont
  {Zhou}}, \bibinfo {author} {\bibfnamefont {K.}~\bibnamefont {Neupane}},
  \bibinfo {author} {\bibfnamefont {Y.~A.}\ \bibnamefont {Nastishin}}, \bibinfo
  {author} {\bibfnamefont {A.~R.}\ \bibnamefont {Baldwin}}, \bibinfo {author}
  {\bibfnamefont {S.~V.}\ \bibnamefont {Shiyanovskii}}, \bibinfo {author}
  {\bibfnamefont {O.~D.}\ \bibnamefont {Lavrentovich}},\ and\ \bibinfo {author}
  {\bibfnamefont {S.}~\bibnamefont {Sprunt}},\ }\bibfield  {title} {\bibinfo
  {title} {Elasticity, viscosity, and orientational fluctuations of a lyotropic
  chromonic nematic liquid crystal disodium cromoglycate},\ }\href
  {https://doi.org/10.1039/C4SM00772G} {\bibfield  {journal} {\bibinfo
  {journal} {Soft Matter}\ }\textbf {\bibinfo {volume} {10}},\ \bibinfo {pages}
  {6571} (\bibinfo {year} {2014})}\BibitemShut {NoStop}%
\bibitem [{\citenamefont {Collings}\ \emph {et~al.}(2017)\citenamefont
  {Collings}, \citenamefont {van~der Asdonk}, \citenamefont {Martinez},
  \citenamefont {Tortora},\ and\ \citenamefont {Kouwer}}]{collings:anchoring}%
  \BibitemOpen
  \bibfield  {author} {\bibinfo {author} {\bibfnamefont {P.~J.}\ \bibnamefont
  {Collings}}, \bibinfo {author} {\bibfnamefont {P.}~\bibnamefont {van~der
  Asdonk}}, \bibinfo {author} {\bibfnamefont {A.}~\bibnamefont {Martinez}},
  \bibinfo {author} {\bibfnamefont {L.}~\bibnamefont {Tortora}},\ and\ \bibinfo
  {author} {\bibfnamefont {P.~H.~J.}\ \bibnamefont {Kouwer}},\ }\bibfield
  {title} {\bibinfo {title} {Anchoring strength measurements of a lyotropic
  chromonic liquid crystal on rubbed polyimide surfaces},\ }\href
  {https://doi.org/10.1080/02678292.2016.1269372} {\bibfield  {journal}
  {\bibinfo  {journal} {Liq. Cryst.}\ }\textbf {\bibinfo {volume} {44}},\
  \bibinfo {pages} {1165} (\bibinfo {year} {2017})}\BibitemShut {NoStop}%
\bibitem [{\citenamefont {Kim}\ \emph {et~al.}(2016)\citenamefont {Kim},
  \citenamefont {Nayani}, \citenamefont {Jeong}, \citenamefont {Jeon},
  \citenamefont {Yoo}, \citenamefont {Lee}, \citenamefont {Park}, \citenamefont
  {Srinivasarao},\ and\ \citenamefont {Jung}}]{kim:macroscopic}%
  \BibitemOpen
  \bibfield  {author} {\bibinfo {author} {\bibfnamefont {J.~Y.}\ \bibnamefont
  {Kim}}, \bibinfo {author} {\bibfnamefont {K.}~\bibnamefont {Nayani}},
  \bibinfo {author} {\bibfnamefont {H.~S.}\ \bibnamefont {Jeong}}, \bibinfo
  {author} {\bibfnamefont {H.-J.}\ \bibnamefont {Jeon}}, \bibinfo {author}
  {\bibfnamefont {H.-W.}\ \bibnamefont {Yoo}}, \bibinfo {author} {\bibfnamefont
  {E.~H.}\ \bibnamefont {Lee}}, \bibinfo {author} {\bibfnamefont {J.~O.}\
  \bibnamefont {Park}}, \bibinfo {author} {\bibfnamefont {M.}~\bibnamefont
  {Srinivasarao}},\ and\ \bibinfo {author} {\bibfnamefont {H.-T.}\ \bibnamefont
  {Jung}},\ }\bibfield  {title} {\bibinfo {title} {Macroscopic alignment of
  chromonic liquid crystals using patterned substrates},\ }\href
  {https://doi.org/10.1039/C5CP07570J} {\bibfield  {journal} {\bibinfo
  {journal} {Phys. Chem. Chem. Phys.}\ }\textbf {\bibinfo {volume} {18}},\
  \bibinfo {pages} {10362} (\bibinfo {year} {2016})}\BibitemShut {NoStop}%
\bibitem [{\citenamefont {Blinov}\ and\ \citenamefont
  {Chigrinov}(1994)}]{blinov:electrooptic}%
  \BibitemOpen
  \bibfield  {author} {\bibinfo {author} {\bibfnamefont {L.~M.}\ \bibnamefont
  {Blinov}}\ and\ \bibinfo {author} {\bibfnamefont {V.~G.}\ \bibnamefont
  {Chigrinov}},\ }\href@noop {} {\emph {\bibinfo {title} {Electrooptic Effects
  in Liquid Crystal Materials}}},\ Partially Ordered Systems\ (\bibinfo
  {publisher} {Springer},\ \bibinfo {address} {New York},\ \bibinfo {year}
  {1994})\BibitemShut {NoStop}%
\bibitem [{\citenamefont {Faetti}\ and\ \citenamefont
  {Marianelli}(2005)}]{faetti:strong}%
  \BibitemOpen
  \bibfield  {author} {\bibinfo {author} {\bibfnamefont {S.}~\bibnamefont
  {Faetti}}\ and\ \bibinfo {author} {\bibfnamefont {P.}~\bibnamefont
  {Marianelli}},\ }\bibfield  {title} {\bibinfo {title} {Strong azimuthal
  anchoring energy at a nematic-polyimide interface},\ }\href
  {https://doi.org/10.1103/PhysRevE.72.051708} {\bibfield  {journal} {\bibinfo
  {journal} {Phys. Rev. E}\ }\textbf {\bibinfo {volume} {72}},\ \bibinfo
  {pages} {051708} (\bibinfo {year} {2005})}\BibitemShut {NoStop}%
\bibitem [{\citenamefont {{S. Faetti}}\ and\ \citenamefont {{G. C.
  Mutinati}}(2003)}]{faetti:improved}%
  \BibitemOpen
  \bibfield  {author} {\bibinfo {author} {\bibnamefont {{S. Faetti}}}\ and\
  \bibinfo {author} {\bibnamefont {{G. C. Mutinati}}},\ }\bibfield  {title}
  {\bibinfo {title} {An improved reflectometric method to measure the azimuthal
  anchoring energy of nematic liquid crystals},\ }\href
  {https://doi.org/10.1140/epje/i2002-10114-1} {\bibfield  {journal} {\bibinfo
  {journal} {Eur. Phys. J. E}\ }\textbf {\bibinfo {volume} {10}},\ \bibinfo
  {pages} {265} (\bibinfo {year} {2003})}\BibitemShut {NoStop}%
\bibitem [{\citenamefont {Virga}(1989)}]{virga:drops}%
  \BibitemOpen
  \bibfield  {author} {\bibinfo {author} {\bibfnamefont {E.~G.}\ \bibnamefont
  {Virga}},\ }\bibfield  {title} {\bibinfo {title} {Drops of nematic liquid
  crystals},\ }\href {https://doi.org/https://doi.org/10.1007/BF00251555}
  {\bibfield  {journal} {\bibinfo  {journal} {Arch. Rational Mech. Anal.}\
  }\textbf {\bibinfo {volume} {107}},\ \bibinfo {pages} {371} (\bibinfo {year}
  {1989})},\ \bibinfo {note} {reprinted in
  \cite{virga:drops_reprinted}}\BibitemShut {NoStop}%
\bibitem [{\citenamefont {Wulff}(1901)}]{wulff:frage}%
  \BibitemOpen
  \bibfield  {author} {\bibinfo {author} {\bibfnamefont {G.}~\bibnamefont
  {Wulff}},\ }\bibfield  {title} {\bibinfo {title} {Zur {F}rage der
  {G}eschwindigkeit des {W}achsthums und der {A}ufl\"osung der
  {K}ristallfl\"achen},\ }\href {https://doi.org/10.1524/zkri.1901.34.1.449}
  {\bibfield  {journal} {\bibinfo  {journal} {Z. Kristallographie und
  Mineralogie}\ }\textbf {\bibinfo {volume} {34}},\ \bibinfo {pages} {449}
  (\bibinfo {year} {1901})}\BibitemShut {NoStop}%
\bibitem [{\citenamefont {Williams}(1986)}]{williams:transitions}%
  \BibitemOpen
  \bibfield  {author} {\bibinfo {author} {\bibfnamefont {R.~D.}\ \bibnamefont
  {Williams}},\ }\bibfield  {title} {\bibinfo {title} {Two transitions in
  tangentially anchored nematic droplets},\ }\href
  {https://doi.org/10.1088/0305-4470/19/16/019} {\bibfield  {journal} {\bibinfo
   {journal} {J. Phys. A: Math. Gen.}\ }\textbf {\bibinfo {volume} {19}},\
  \bibinfo {pages} {3211} (\bibinfo {year} {1986})}\BibitemShut {NoStop}%
\bibitem [{\citenamefont {Kim}\ \emph {et~al.}(2013)\citenamefont {Kim},
  \citenamefont {Shiyanovskii},\ and\ \citenamefont
  {Lavrentovich}}]{kim:morphogenesis}%
  \BibitemOpen
  \bibfield  {author} {\bibinfo {author} {\bibfnamefont {Y.-K.}\ \bibnamefont
  {Kim}}, \bibinfo {author} {\bibfnamefont {S.~V.}\ \bibnamefont
  {Shiyanovskii}},\ and\ \bibinfo {author} {\bibfnamefont {O.~D.}\ \bibnamefont
  {Lavrentovich}},\ }\bibfield  {title} {\bibinfo {title} {Morphogenesis of
  defects and tactoids during isotropic-nematic phase transition in
  self-assembled lyotropic chromonic liquid crystals},\ }\href@noop {}
  {\bibfield  {journal} {\bibinfo  {journal} {J. Phys.: Condens. Matter}\
  }\textbf {\bibinfo {volume} {25}},\ \bibinfo {pages} {404202} (\bibinfo
  {year} {2013})}\BibitemShut {NoStop}%
\bibitem [{\citenamefont {McGinn}\ \emph {et~al.}(2013)\citenamefont {McGinn},
  \citenamefont {Laderman}, \citenamefont {Zimmermann}, \citenamefont
  {Kitzerow},\ and\ \citenamefont {Collings}}]{mcginn:planar}%
  \BibitemOpen
  \bibfield  {author} {\bibinfo {author} {\bibfnamefont {C.~K.}\ \bibnamefont
  {McGinn}}, \bibinfo {author} {\bibfnamefont {L.~I.}\ \bibnamefont
  {Laderman}}, \bibinfo {author} {\bibfnamefont {N.}~\bibnamefont
  {Zimmermann}}, \bibinfo {author} {\bibfnamefont {H.-S.}\ \bibnamefont
  {Kitzerow}},\ and\ \bibinfo {author} {\bibfnamefont {P.~J.}\ \bibnamefont
  {Collings}},\ }\bibfield  {title} {\bibinfo {title} {Planar anchoring
  strength and pitch measurements in achiral and chiral chromonic liquid
  crystals using 90-degree twist cells},\ }\href
  {https://doi.org/10.1103/PhysRevE.88.062513} {\bibfield  {journal} {\bibinfo
  {journal} {Phys. Rev. E}\ }\textbf {\bibinfo {volume} {88}},\ \bibinfo
  {pages} {062513} (\bibinfo {year} {2013})}\BibitemShut {NoStop}%
\bibitem [{\citenamefont {McIntyre}\ and\ \citenamefont
  {Snyder}(1978)}]{mcintyre:light}%
  \BibitemOpen
  \bibfield  {author} {\bibinfo {author} {\bibfnamefont {P.}~\bibnamefont
  {McIntyre}}\ and\ \bibinfo {author} {\bibfnamefont {A.~W.}\ \bibnamefont
  {Snyder}},\ }\bibfield  {title} {\bibinfo {title} {Light propagation in
  twisted anisotropic media: {A}pplication to photoreceptors},\ }\href
  {https://doi.org/10.1364/JOSA.68.000149} {\bibfield  {journal} {\bibinfo
  {journal} {J. Opt. Soc. Am.}\ }\textbf {\bibinfo {volume} {68}},\ \bibinfo
  {pages} {149} (\bibinfo {year} {1978})}\BibitemShut {NoStop}%
\bibitem [{\citenamefont {McIntyre}(1978)}]{mcintyre:transmission}%
  \BibitemOpen
  \bibfield  {author} {\bibinfo {author} {\bibfnamefont {P.}~\bibnamefont
  {McIntyre}},\ }\bibfield  {title} {\bibinfo {title} {Transmission of light
  through a twisted nematic liquid-crystal layer},\ }\href
  {https://doi.org/10.1364/JOSA.68.000869} {\bibfield  {journal} {\bibinfo
  {journal} {J. Opt. Soc. Am.}\ }\textbf {\bibinfo {volume} {68}},\ \bibinfo
  {pages} {869} (\bibinfo {year} {1978})}\BibitemShut {NoStop}%
\bibitem [{\citenamefont {Mauguin}(1911)}]{mauguin:representation}%
  \BibitemOpen
  \bibfield  {author} {\bibinfo {author} {\bibfnamefont {C.}~\bibnamefont
  {Mauguin}},\ }\bibfield  {title} {\bibinfo {title} {Sur la repr\'sentation
  g\'eometrique de poincar\'e relative aux propri\'et\'es optiques des piles de
  lames},\ }\href@noop {} {\bibfield  {journal} {\bibinfo  {journal} {Bull.
  Soc. Fr. Min\'er.}\ }\textbf {\bibinfo {volume} {34}},\ \bibinfo {pages} {6}
  (\bibinfo {year} {1911})},\ \bibinfo {note} {available from
  \url{https://gallica.bnf.fr/ark:/12148/bpt6k1089141/f23.item}}\BibitemShut
  {NoStop}%
\bibitem [{\citenamefont {Paparini}\ and\ \citenamefont
  {Virga}(2021{\natexlab{b}})}]{paparini:nematic}%
  \BibitemOpen
  \bibfield  {author} {\bibinfo {author} {\bibfnamefont {S.}~\bibnamefont
  {Paparini}}\ and\ \bibinfo {author} {\bibfnamefont {E.~G.}\ \bibnamefont
  {Virga}},\ }\bibfield  {title} {\bibinfo {title} {Nematic tactoid
  population},\ }\href {https://doi.org/10.1103/PhysRevE.103.022707} {\bibfield
   {journal} {\bibinfo  {journal} {Phys. Rev. E}\ }\textbf {\bibinfo {volume}
  {103}},\ \bibinfo {pages} {022707} (\bibinfo {year}
  {2021}{\natexlab{b}})}\BibitemShut {NoStop}%
\bibitem [{\citenamefont {Ericksen}(1991)}]{ericksen:liquid}%
  \BibitemOpen
  \bibfield  {author} {\bibinfo {author} {\bibfnamefont {J.~L.}\ \bibnamefont
  {Ericksen}},\ }\bibfield  {title} {\bibinfo {title} {Liquid crystals with
  variable degree of orientations},\ }\href
  {https://doi.org/10.1007/BF00380413} {\bibfield  {journal} {\bibinfo
  {journal} {Arch. Rational Mech. Anal.}\ }\textbf {\bibinfo {volume} {113}},\
  \bibinfo {pages} {97} (\bibinfo {year} {1991})}\BibitemShut {NoStop}%
\bibitem [{\citenamefont {Poggi}\ \emph {et~al.}(1976)\citenamefont {Poggi},
  \citenamefont {Filippini},\ and\ \citenamefont {Aleonard}}]{poggi:free}%
  \BibitemOpen
  \bibfield  {author} {\bibinfo {author} {\bibfnamefont {Y.}~\bibnamefont
  {Poggi}}, \bibinfo {author} {\bibfnamefont {J.}~\bibnamefont {Filippini}},\
  and\ \bibinfo {author} {\bibfnamefont {R.}~\bibnamefont {Aleonard}},\
  }\bibfield  {title} {\bibinfo {title} {The free energy as a function of the
  order parameter in nematic liquid crystals},\ }\href
  {https://doi.org/https://doi.org/10.1016/0375-9601(76)90452-7} {\bibfield
  {journal} {\bibinfo  {journal} {Phys. Lett. A}\ }\textbf {\bibinfo {volume}
  {57}},\ \bibinfo {pages} {53} (\bibinfo {year} {1976})}\BibitemShut {NoStop}%
\bibitem [{\citenamefont {Davidson}\ \emph
  {et~al.}(2015{\natexlab{b}})\citenamefont {Davidson}, \citenamefont {Kang},
  \citenamefont {Jeong}, \citenamefont {Still}, \citenamefont {Collings},
  \citenamefont {Lubensky},\ and\ \citenamefont {Yodh}}]{davidson:erratum}%
  \BibitemOpen
  \bibfield  {author} {\bibinfo {author} {\bibfnamefont {Z.~S.}\ \bibnamefont
  {Davidson}}, \bibinfo {author} {\bibfnamefont {L.}~\bibnamefont {Kang}},
  \bibinfo {author} {\bibfnamefont {J.}~\bibnamefont {Jeong}}, \bibinfo
  {author} {\bibfnamefont {T.}~\bibnamefont {Still}}, \bibinfo {author}
  {\bibfnamefont {P.~J.}\ \bibnamefont {Collings}}, \bibinfo {author}
  {\bibfnamefont {T.~C.}\ \bibnamefont {Lubensky}},\ and\ \bibinfo {author}
  {\bibfnamefont {A.~G.}\ \bibnamefont {Yodh}},\ }\bibfield  {title} {\bibinfo
  {title} {Erratum: {C}hiral structures and defects of lyotropic chromonic
  liquid crystals induced by saddle-splay elasticity [{P}hys. {R}ev. {E} 91,
  050501({R}) (2015)]},\ }\href {https://doi.org/10.1103/PhysRevE.92.019905}
  {\bibfield  {journal} {\bibinfo  {journal} {Phys. Rev. E}\ }\textbf {\bibinfo
  {volume} {92}},\ \bibinfo {pages} {019905} (\bibinfo {year}
  {2015}{\natexlab{b}})}\BibitemShut {NoStop}%
\bibitem [{\citenamefont {Virga}(1991)}]{virga:drops_reprinted}%
  \BibitemOpen
  \bibfield  {author} {\bibinfo {author} {\bibfnamefont {E.~G.}\ \bibnamefont
  {Virga}},\ }\bibfield  {title} {\bibinfo {title} {Drops of nematic liquid
  crystals},\ }in\ \href@noop {} {\emph {\bibinfo {booktitle} {Mechanics and
  Thermodynamics of Continua}}},\ \bibinfo {editor} {edited by\ \bibinfo
  {editor} {\bibfnamefont {H.}~\bibnamefont {Markovitz}}, \bibinfo {editor}
  {\bibfnamefont {V.~J.}\ \bibnamefont {Mizel}},\ and\ \bibinfo {editor}
  {\bibfnamefont {D.~R.}\ \bibnamefont {Owen}}}\ (\bibinfo  {publisher}
  {Springer},\ \bibinfo {address} {Berlin},\ \bibinfo {year} {1991})\ pp.\
  \bibinfo {pages} {211--230}\BibitemShut {NoStop}%
\end{thebibliography}
%

\end{document}